\documentclass[twocolumn]{aastex62}
\graphicspath{{./}{figures/}}

\received{}
\revised{}
\accepted{}
\submitjournal{ApJ}

\shorttitle{Radial variations in grain sizes and dust scale heights}
\shortauthors{Ohashi \& Kataoka}

\begin{document}

\title{Radial variations in grain sizes and dust scale heights in the protoplanetary disk around HD 163296 revealed by ALMA polarization observation}

\author{Satoshi Ohashi}
\affil{RIKEN Cluster for Pioneering Research, 2-1, Hirosawa, Wako-shi, Saitama 351-0198, Japan}
\email{satoshi.ohashi@riken.jp}

\author{Akimasa Kataoka}
\affiliation{National Astronomical Observatory of Japan, 2-21-1 Osawa, Mitaka, Tokyo 181-8588, Japan}



\begin{abstract}
The HD 163296 disk shows ring and gap substructures with ALMA observations.
In addition, this is the only disk where the rings and gaps are spatially resolved in millimeter-wave polarization measurements.
In this paper, we conduct radiative transfer modeling that includes self-scattering polarization to constrain the grain size and its distribution. We found that the grain size and dust scale height are the key parameters for reproducing the radial and azimuthal distributions of the observed polarization signature.
Radial variation is mainly determined by grain size. The polarization fraction is high if the particle size is $\sim \lambda/2\pi$; it is low if the particle size is larger or smaller than this.
In contrast, azimuthal variation in polarization is enhanced if the dust scale height is increased.
Based on detailed modeling of the HD 163296 polarization, we found the following radial variations in the grain size and dust scale height. The maximum grain size was 140 microns  in the gaps and significantly larger or smaller in the rings.
The dust scale height is less than one-third the gas scale height inside the 70 au  ring, and two-thirds the gas scale height outside the 70 au ring.
Furthermore, we constrained the gas turbulence to be $\alpha\lesssim1.5\times10^{-3}$ in the 50 au gap and $\alpha\sim 0.015-0.3$ in the 90 au gap.
The transition of the turbulence strength at the boundary of the 70 au ring indicates the existence of a dead zone.
\end{abstract}



\section{Introduction} \label{sec:intro}

Millimeter and sub-millimeter polarization has been detected from more than ten protoplanetary disks thanks to the high spatial resolution and high sensitivity of the Atacama Large Mmillimeter/submillimeter Array (ALMA) \citep[e.g.,][]{kat16,ste17,lee18,cox18,hul18,sad18,har18,alv18,sad19}.
However, the mechanism causing the polarization is not a straightforward extension of that in star-forming regions, where the polarization is believed to be emitted from elongated dust grains aligned with magnetic fields \citep[e.g.,][]{laz07,and15}.
This polarization is used as a tracer of magnetic field directions \citep[e.g.,][]{rao98,lai01,gir06,ste13,cox15,hul17,mau18,tak18,kwo19,gou19}.
Multiple origins have been proposed to explain the millimeter polarization by protoplanetary disks, including grain alignment with magnetic fields \citep{cho07,ber17}, radiation fields \citep{laz07b,taz17}, or gas flow \citep{kat19}, and the self-scattering of thermal dust emission \citep{kat15,yang16}.

The polarization mechanism strongly depends on the grain size. 
\citet{taz17} theoretically showed that dust grains smaller than $\sim100$ microns are aligned with the direction of magnetic fields whereas those larger than $\sim100$ microns are aligned with the direction of radiation fields; the threshold of 100 microns depends on the properties of the dust grains and the strength of the magnetic field. 
Note that \citet{yang19,kat19} also found that grains may be aligned with gas flow, but \citet{taz17} did not report this effect.
\citet{kat15} found that the self-scattering of thermal dust emission can be detected only when the grain radius is around $\sim\lambda/2\pi$, where $\lambda$ is the observation wavelength.
Considering the significant diversity in grain growth in protoplanetary disks, we expect that the mechanisms of polarization will be different for different disks and will depend on the location in the disks.

Previous observations have shown that the polarization mechanism depends on location for a lopsided disk. 
\citet{kat16} found a sharp change in the direction of the polarization vectors, from the radial direction to the azimuthal direction, in the north region of the HD 142527 disk, which is consistent with self-scattering of in a face-on disk.
\citet{oha18} further analyzed the same target and reporterd that the lopsided protoplanetary disk of HD 142527 has polarization caused by self-scattering in the northern region and that caused by grain alignment with toroidal magnetic fields in the southern region at a 0.87 mm wavelength.
This indicates that the grain size is smaller in the southern region and larger in the northern region, which is consistent with the azimuthal dust trapping scenario where dust grains are trapped at the gas pressure bump azimuthally \citep[e.g.,][]{van13,fuk13,per14,pin15,cas15,boe17}.

In contrast, inclined smooth disks usually show self-scattering features. 
Several disks have polarization vectors parallel to the disk minor axis \citep[e.g.,][]{ste14,kat17,ste17,lee18,cox18,gir18,sad18,hul18,har18,bac18,den19,har19,mor19}, which is consistent with the self-scattering model for an inclined disk \citep{yang16,yang17,kat16}.
Note that HL Tau shows a self-scattering feature in Band 7 but azimuthal polarization in Band 3, which can be interpreted as radiative alignment or gas flow alignment \citep{kat17,yang19}.

In particular, polarization of the protoplanetary disk of HD 163296 was observed with a spatial resolution  sufficient to resolve ring and gap structures \citep{den19}.
It was found that the polarization vectors are basically parallel to the disk minor axis in the central region; however, there are additional azimuthal components in gaps.
Furthermore, the polarization fraction has a peak in the gaps between dust rings.
These radial and azimuthal variations of the polarization fraction may reflect the dust properties and distributions in the protoplanetary disk.
In this study, we perform radiative transfer modeling to determine what can be constrained based on spatially resolved polarimetric observations.
We then model the continuum and polarized emissions of the HD 163296 disk to extract information on dust grains.

The protoplanetary disk around the Herbig Ae star HD 163296 has been well studied not only with ALMA \citep[e.g.,][]{de13,ros13,fla15,ise16,ise18,not19} but also with optical and infrared telescopes such as the Hubble Space Telescope ({\it HST}), the Very Large Telescope ({\it VLT}), and {\it SPHERE} (Spectro-Polarimetric High-contrast Exoplanet REsearch instrument) \citep[e.g.,][]{gra00,kra08,mur18}.
The distance was derived to be 101.5 pc by \citet{gai18,bai18}.
The ALMA DSHARP project \citep{and18} found several rings using a high angular resolution of $\sim3$ au and reported the detailed azimuthal asymmetries in the disk \citep{ise18,hua18}. These rings and asymmetries may be produced by planets \citep{zha18,liu18}.
The presence of planets is also suggested by deviations from Keplerian rotational motions \citep{tea18,pin18}.
However, optical and infrared high-contrast imaging has not identified planetary mass companions yet, but has set the upper limits for the masses of giant planets of a few $M_J$ \citep{gui18,mes19}.

 Ring formation mechanism is still under discussion even though ring structures have been discovered in many protoplanetary disks with ALMA \citep[e.g.,][]{alma15,and16,can16,ise16,tsu16,cie17,loo17,kra17,van17,and18,ans18,ber18,boe18,cla18,dip18,don18,fed18,lon18,she18,van18,van19}.
However, the ring formation is still under discussion.
The presence of planets in a gap in the disk around PDS 70 was detected based on the H$\alpha$ emission of accreting gas from protoplanets \citep{kep18,haf19}. Furthermore, the presence of asymmetries may also indicate planets or stellar companions.
However, besides planets, dust opacity variations at snow lines of several molecular species \citep{zha15,oku16}, secular gravitational instability \citep{tak14,tom18,tom19}, dust accumulation at the edge of a `dead zone' \citep{flo15,rug16}, and other mechanisms \citep{lor15,bet16,dul18b} be create ring and gap structures.
 A recent sample study found no correlation between gap radii and disk temperature, suggesting that a snow line model is unlikely to be a common origin of multiring systems \citep{van19}.

The rest of this paper is organized as follows.
In Section \ref{sec:data}, we analyze the polarization data of HD 163296 and describe the observed features in the disk.
Section \ref{sec:model} explains the method used to obtain the disk model and the radiative transfer calculations.
In Section \ref{sec:scat}, we investigate the polarization patterns of self-scattering in terms of grain size and dust scale height based on a smoothed disk without any rings as a simple case.
In Section \ref{sec:target}, we apply the self-scattering model to the HD 163296 disk and investigate the distributions of grain size and dust scale height.
In Section \ref{sec:discussion}, we discuss the disk structure by considering the observed variations of grain size and dust scale height. We estimate the turbulence strength $\alpha$ and discuss the existence of a dead zone.
Finally, the results and the conclusions of this work are summarized in Section \ref{sec:conclusion}.

\section{ALMA Polarization Data of HD 163296} \label{sec:data}

\begin{figure*}[htbp]
  \begin{center}
  \includegraphics[width=18.cm,bb=0 0 2833 821]{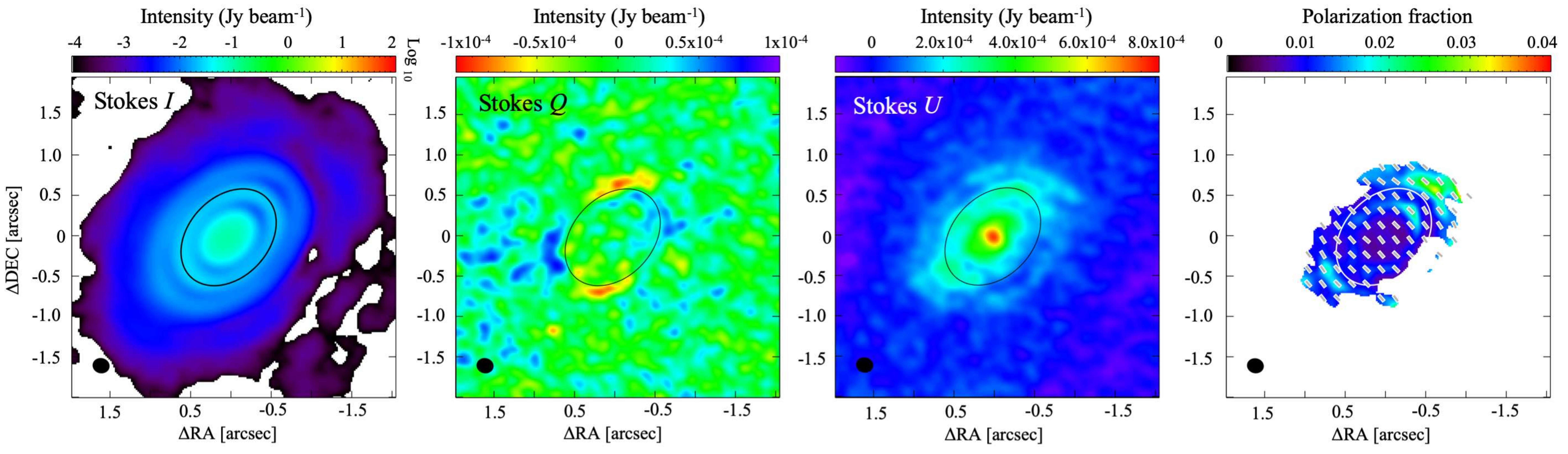}
  \end{center}
  \caption{0.87 mm dust continuum polarization in HD 163296. The panels show the total intensity (Stokes {\it I}), Stokes {\it Q} component, Stokes {\it U} component, and polarization fraction, respectively. The black and white circles indicate the position of the 70 au ring. The polarization vectors are overlaid on the polarization fraction. The polarization vectors and the polarization fraction are shown where the polarized intensity is higher than $3\sigma_{\rm PI}$. Note that the lengths of the polarization vectors are set to be the same.
  }
  \label{obs}
\end{figure*}

\subsection{Analysis of Archival Data}

We used the archival data of the 0.87 mm dust polarization from the protoplanetary disk around HD 163296 (2015.1.00616.S; PI: C. Pinte).
These data were originally presented by \citet{den19}.
The reduction and calibration of the data were done with CASA version 4.5.3 \citep{mcm07} in a standard manner.

Stokes images were reconstructed using the CASA task tCLEAN, with Briggs weighting with a robust parameter of 0.5. The pixel size was set to $0\farcs02$.
In addition, to improve sensitivity and image fidelity, self-calibration for both phase and amplitude was performed.
The beam size of the final product was $0\farcs20\times0\farcs18$, corresponding to a spatial resolution of $\sim20\times18$ au at the assumed distance of 101.5 pc.
Note that \citet{den19} examined the calibrated data in more detail, and applied further flagging. However, we found that our simple calibration is consistent with the results of \citet{den19}, as shown in the following section.

Stokes {\it Q} and {\it U} components produce polarized intensity ($PI$). Note that we ignore the Stokes {\it V} component in this study because it has not been well characterized for ALMA. 
The $PI$ value has a positive bias because it is always a positive quantity. This bias has a particularly significant effect in low-signal-to-noise measurements. We thus debiased the polarized intensity map as
$PI=\sqrt{Q^2+U^2-\sigma_{\rm PI}^2}$.
The root-mean-square (rms) noise of the Stokes {\it Q}, and {\it U}, and the polarized intensity was derived to be $\sigma_{\it Q}=1.7\times10^{-5}$ Jy beam$^{-1}$, $\sigma_{\it U}=3.0\times10^{-5}$ Jy beam$^{-1}$, and $\sigma_{\rm PI}=4.0\times10^{-5}$ Jy beam$^{-1}$, respectively. 
 The rms values of the Stokes {\it Q} and {\it U} emissions were calculated from an area in the images in which emission was not detected.
The polarization fraction ($P_{\rm frac}=PI/I$) was derived only where detection was above the threshold 3$\sigma_{\rm PI}$.

\subsection{Results} \label{sec:obs}

Figure \ref{obs} shows images of the Stokes {\it I}, {\it Q}, and {\it U}, and the polarization fraction of the protoplanetary disk around HD 163296 at a 0.87 mm wavelength.
The polarization vectors are overlaid on the image of the polarization fraction.
The location of the continuum ring at a radius of 70 au is shown by a black or white circle.
These images are consistent with those obtained by \citet{den19}.

The polarization fraction is $\sim0.8\pm0.04$\% in the center and increases to $\sim2.3\pm0.5$\% in the first gap at 50 au and to $\sim3\pm0.7$\% in the second gap at 90 au along the major axis.
Polarization fraction along the minor axis is lower than that along the major axis.
The polarized intensity outside the 70 au ring was not detected along the minor axis.

The polarization vectors are mainly aligned with the disk minor axis.
In addition, we confirmed the azimuthal twist of $\pm10^\circ$ in the polarization vectors reported by \citet{den19}. This twist is located outside the 70 au ring and can be recognized in the Stokes {\it Q} image because the Stokes {\it Q} emission represents polarization pointing in the disk major axis in this disk geometry.

In this paper, we perform radiative transfer calculations by taking into account the polarization due to only self-scattering.
This assumption is consistent with the results reported by \citet{den19}, who indicated that the polarization is produced by the self-scattering without any additional dust alignment mechanism.

\section{Disk Model}\label{sec:model}

To perform radiative transfer calculations, we constructed an axisymmetric dust disk model with several additional rings that reproduce the continuum image. We discuss the physical parameters of the disk in the following section.

We adopted the following temperature profile with a smooth power law distribution derived by \citet{ise16}:
\begin{equation}
T_{d}= 220\ {\rm K}\ \Big(\frac{R}{1\ {\rm au}}\Big)^{-0.5}. 
\label{temp}
\end{equation}

We adopted a power-law surface density profile with an exponential tail and added exponential bright regions to mimic the observation profile.
We took the intensity profile to be on the major axis in the northwest direction to avoid the additional crescent-shaped structure reported in the opposite direction \citep{ise18}.
The intensity can be expressed as
\begin{equation}
    I_{\nu}(R)=(1-e^{-\tau(R)})B_{\nu}(T_{d}(R)).
\label{in}
\end{equation}
Using this equation, we assume the radial optical depth profile as to be
\begin{eqnarray}
\tau=&0.14\Big(\frac{R}{1\ {\rm au}}\Big)^{0.5}\exp\Big(-\Big(\frac{R}{35\ {\rm au}}\Big)^5\Big)\nonumber\\
&+1.1\exp\Big(-\frac{1}{2}\Big(\frac{R-70{\rm au}}{3.8\ {\rm au}}\Big)^2\Big)\nonumber\\
&+0.2\exp\Big(-\frac{1}{2}\Big(\frac{R-104{\rm au}}{5.6\ {\rm au}}\Big)^2\Big)\nonumber\\
&+0.03\exp\Big(-\frac{1}{2}\Big(\frac{R-160{\rm au}}{16\ {\rm au}}\Big)^2\Big).
\label{tau}
\end{eqnarray}
The surface density is derived with the assumption that $\tau=\kappa_{\rm abs}\times\Sigma_{\rm dust}$, where $\kappa_{\rm abs}$ is the absorption opacity and $\Sigma_{\rm dust}$ is the dust surface density.
 The radial profile of the intensity is set using the equations \ref{temp}, \ref{in}, and \ref{tau}.
The unit of intensity is set to be Jy beam$^{-1}$ for comparison with the observations.

The calculation of opacity followed that by \citet{kat16}.
The dust grains were assumed to be spherical and to have a power-law size distribution with an exponent of $q=-3.5$ \citep{mat77} and maximum grain size $a_{\rm max}$.
This maximum grain size is considered to be the representative grain size in the following discussion. The opacity was calculated using Mie theory. 
The composition was assumed to be a mixture of silicate (50\%) and water ice (50\%) \citep{pol94,kat14}. We used the refractive index of astronomical silicate \citep{wei01} and water ice \citep{war84} and calculated their mixture based on effective medium theory using the Maxwell-Garnett rule \cite[e.g.,][]{boh83,miy93}. 
Note that a different abundance may lead to a different absolute value of polarization degree, which should be investigated in future studies.

\begin{figure}[htbp]
  \begin{center}
  \includegraphics[width=8.4cm,bb=0 0 1963 1289]{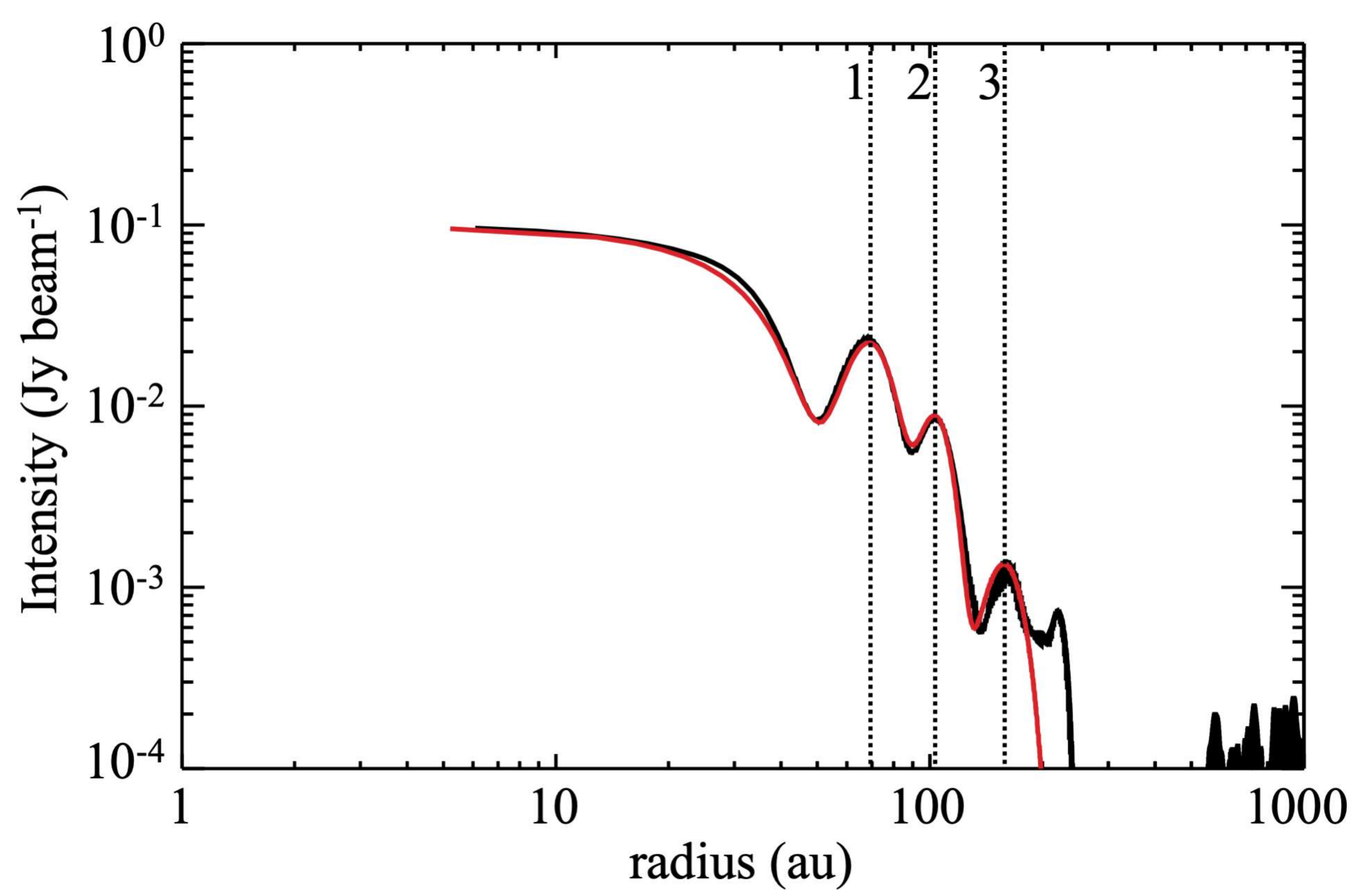}
  \end{center}
  \caption{Intensity profile for dust continuum observation along the major axis in the North-West direction (black line) and intensity profile for the disk model calculated using RADMC-3D (red line). The radii of the three intensity rings are illustrated by the dotted vertical lines.
  }
  \label{radial}
\end{figure}

The vertical density distribution is assumed to be Gaussian with a dust scale height $h_d$ such that $\rho_d=\Sigma_d/(\sqrt{2\pi}h_d)\exp(-z^2/h_d^2)$.
We set an additional dust settling parameter $f_{\rm set}$ such that $h_d=h_g/f_{\rm set}$, where $h_g$ is the gas scale height ($h_g=c_s/\Omega$), to mimic grain settling.

We performed radiative transfer calculations with RADMC-3D\footnote{RADMC-3D is an open code of radiative transfer calculations developed by Cornelis Dullemond. The code is available online at: \url{http://www.ita.uniheidelberg.de/~dullemond/software/radmc-3d/}} \citep{dul12} taking into account multiple scattering, as done by \citet{kat15}.
The inclination and position angle were assumed to be $47^\circ$ and 133.3$^\circ$, respectively, as derived by the DSHARP project \citep{ise18}.

The Stokes {\it I}, {\it Q}, and {\it U} images were convolved with a Gaussian beam with a full width at half maximum (FWHM) of $0.2\arcsec$ to produce the final model images.
Figure \ref{radial} shows the intensity profile (Stokes {\it I}) obtained from the continuum data on the major axis overlaid with the model intensity profile.
The figure confirms that the model reproduces the continuum image.

The three rings are located at 70, 104, and 160 au, respectively.
The width of the rings is 3.8, 5.6, and 16 au, respectively.
\citet{ise18} derived the locations of the two inner rings as 67 au (width: 6.6 au) and 101 au (width: 5.8 au), respectively.
These values are similar to those for our model; the difference is a few au, which is much smaller than the spatial resolution of $\sim20$ au.

\section{Simple Disk Model} \label{sec:scat}

In this section, we first review the basic morphology of polarization vectors in inclined disks with a schematic illustration.
Then, we perform radiative transfer calculations, focusing on the effects of grain size and dust scale height distribution, which are the key parameters in the following modeling of the HD 163296 disk.

\subsection{Polarization Morphology of Inclined Disk}

\begin{figure}[htbp]
  \begin{center}
  \includegraphics[width=8.5cm,bb=0 0 2400 826]{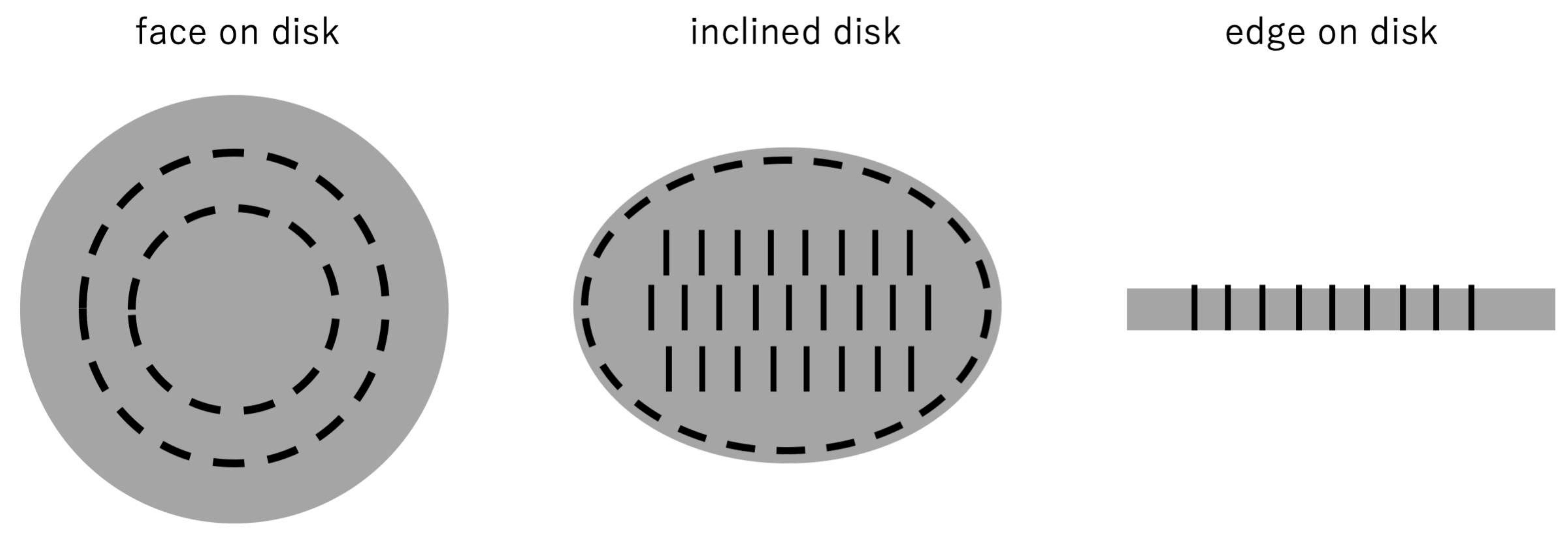}
  \end{center}
  \caption{Schematic views of polarization vectors predicted by self-scattering for face-on, inclined, and edge-on disks. The intensity profile of the disk is assumed to be smooth and axisymmetric.
  }
  \label{view}
\end{figure}

Figure \ref{view} shows schematic illustrations of the polarization vectors for face-on, inclined, and edge-on disks \citep[e.g.,][]{kat16,yang16}. 
The intensity profile is assumed to be smooth and axisymmetric.
The polarization vectors are in the azimuthal directions for the face-on disk.
This is because the radiation gradients are in the radial directions. 
In contrast, the edge-on disk has polarization vectors parallel to the disk minor axis because the dominant radiation is parallel to the disk midplane.
The inclined disk has a combination of the polarization patterns of the face-on and edge-on disks. 
In the central region, the polarization vectors are mostly parallel to the minor axis; those in the azimuthal directions are around the edge of the disk.
With the limited sensitivity of polarization observations, the polarized flux is mainly detected from the central region of inclined disks.
Therefore, polarization vectors parallel to a disk minor axis have been observed in the central part of inclined disks \citep[e.g.,][]{kat16,ste17,hul18,bac18}.
In particular, \citet{bac18} detected the polarization vectors parallel to the disk minor axis in the center and those in the azimuthal directions outside of the disk of DG Tau using high sensitivity observations.
These studies observationally demonstrate that the polarization pattern changes radially in an inclined disk.

\subsection{Dust Grain Size and Dust Scale Height Distribution}

In this subsection, we discuss how dust settling can change the polarization pattern.
To investigate the dependence of the polarization pattern and the polarization fraction on grain size and dust scale height, we performed radiative transfer calculations.
For simplicity, we considered a smooth inclined disk.
The density and optical depth profile were assumed to be
\begin{equation}
\tau=0.14\Big(\frac{R}{1\ {\rm au}}\Big)^{0.5}\exp\Big(-\Big(\frac{R}{35\ {\rm au}}\Big)^5\Big).
\end{equation}
The profile is the same as that in Figure \ref{radial} but without the ring structures.

\begin{figure}[htbp]
  \begin{center}
  \includegraphics[width=8.5cm,bb=0 0 1879 1164]{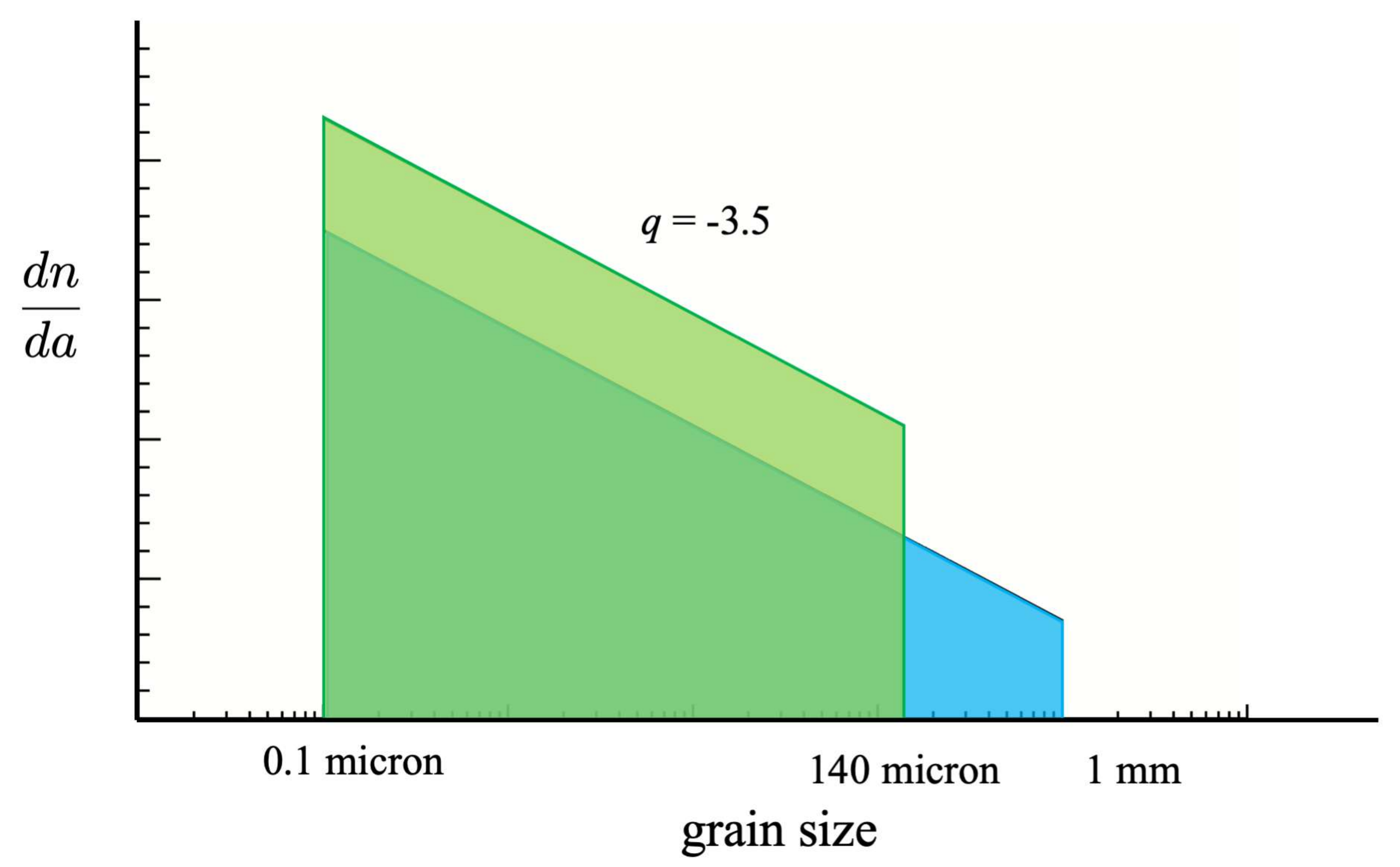}
  \end{center}
  \caption{Number distribution of dust grains as a function of grain size for the two different populations.
  }
  \label{grain_population}
\end{figure}

To investigate effects of grain size distribution, we considered two different grain populations. One consists of dust grains with a power-law size distribution and a maximum grain size of 140 $\mu$m and the other consists of those with a maximum grain size of 1 mm (Figure \ref{grain_population}).  As shown in the figure \ref{grain_population}, the number of dust grains is dominated by the smallest dust grains. However, the minimum grain size is small enough not to affect the results.
The two populations of dust grains are representative of those that contribute to polarization and those do not, respectively: the 140 $\mu$m population produces the most efficient polarization at a 0.87 mm observation wavelength whereas the 1 mm population does not contribute to the polarization, but does to the Stokes {\it I} emission.
To investigate the effects of dust population, we changed the mass ratio of the 140 $\mu$m and 1 mm populations while keeping the total optical depth fixed. 
With this setting, we expect that the 1 mm population will contribute to reduce the polarization fraction without changing the Stokes {\it I} intensity.

\begin{figure*}[htbp]
  \begin{center}
  \includegraphics[width=18cm,bb=0 0 1655 1650]{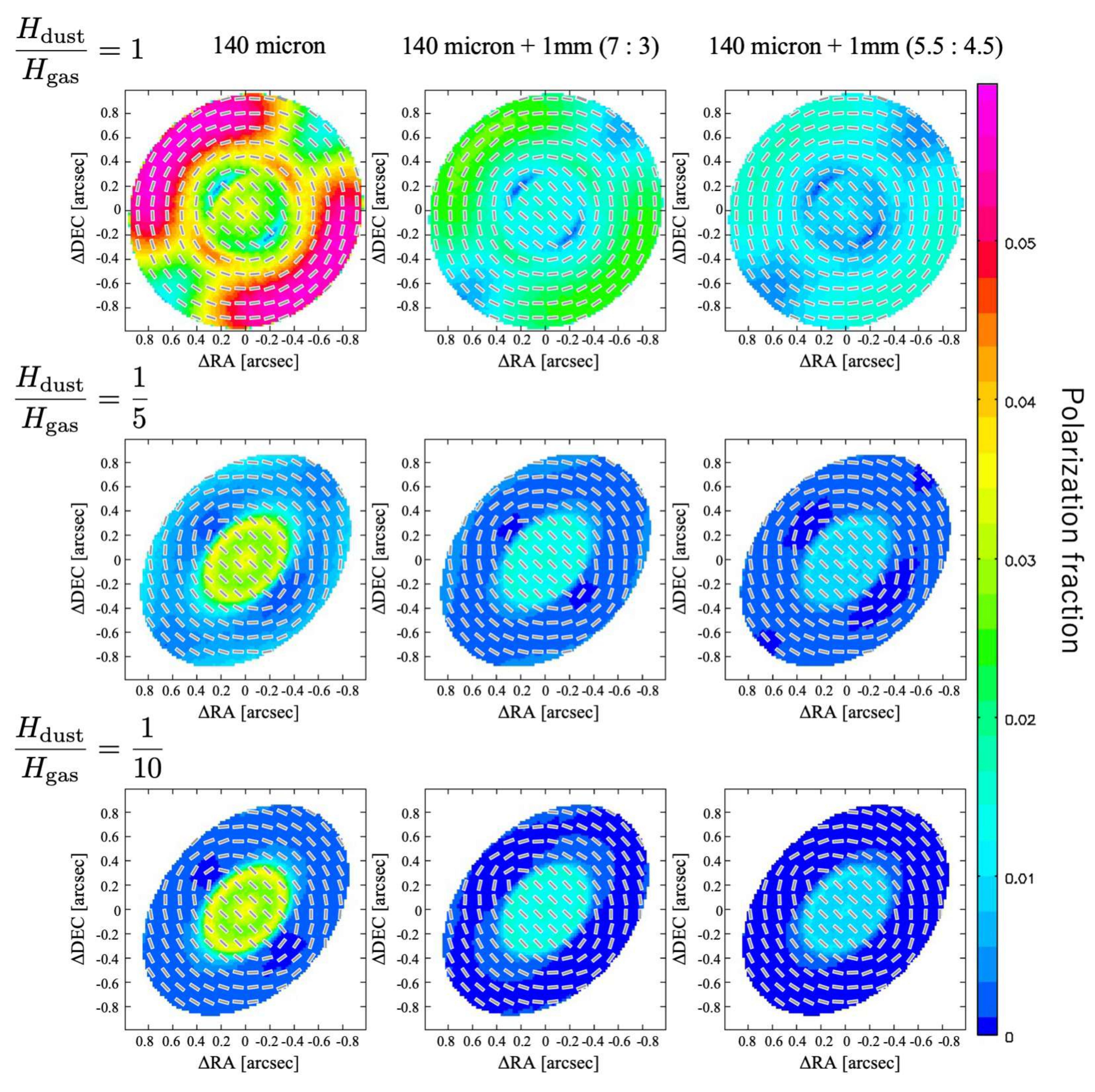}
  \end{center}
  \caption{Polarization fraction overlaid with the polarization vectors calculated using RADMC-3D for various dust grain populations and dust scale heights. The intensity profile of the disk is the same as that for the disk of HD 163296 without the rings.
  }
  \label{fig1}
\end{figure*}

We also investigated the effects of dust scale height distribution.
We performed radiative transfer calculations for various $f_{\rm set}$ values to investigate the effects of dust settling.
Figure \ref{fig1} shows the results obtained for various dust scale heights and dust grain populations.
The left, middle, and right columns show the cases for the 140 $\mu$m population, and combinations of the 140 $\mu$m and 1 mm populations with mass ratios of $7:3$ and $5.5:4.5$, respectively.

Figure \ref{fig1} clearly shows the dependence of the polarization fraction and vectors on dust grain population and dust scale height.
The ratio of the 140 $\mu$m and 1 mm populations changes the polarization fraction but does not change the polarization vectors.
With increasing dust scale height, the azimuthal component is enhanced, especially in the outer region.
As a result, the polarization fraction increases along the minor axis in the outer region.

\begin{figure}[htbp]
  \begin{center}
  \includegraphics[width=8.5cm,bb=0 0 2642 1330]{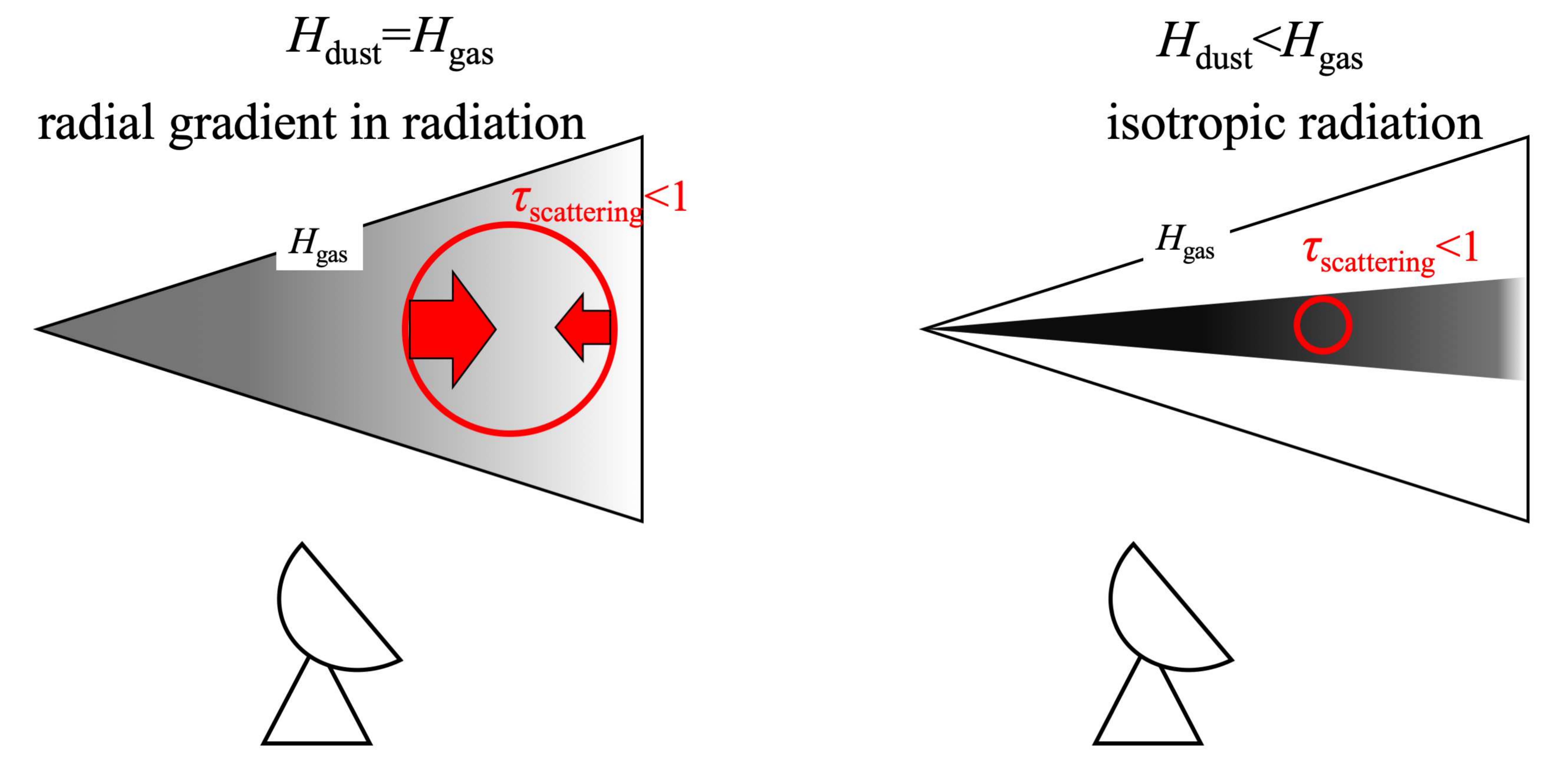}
  \end{center}
  \caption{Schematic views of anisotropic radiation for different dust scale heights. When the dust scale height is increased, the radiation gradient in the radial direction increases.
  }
  \label{scaleheight}
\end{figure}

The enhancement of the azimuthal components with larger dust scale height can be understood by considering the anisotropies of the radiation fields.
Let us consider an increase in dust scale height with the total optical depth and column density fixed.
As shown in Figure \ref{scaleheight}, although the optical depth along the line of sight does not change when the dust scale height increases, the optical depth perpendicular to the line of sight decreases because the dust spatial density decreases. 
Therefore, radiation from farther distances also contributes to the scattering. 
Thus, the incoming flux to be scattered by dust grains has more flux in the radial direction and becomes more anisotropic with increasing dust scale height.
This explains the higher polarization fraction with azimuthal vectors for the thicker disk.

This enhancement of the azimuthal components in the disk with larger dust scale height results in the azimuthal variation of the polarization fraction.
Along the minor axis, we identified a drop of the polarization fraction in the region where the polarization vectors change from the disk minor axis directions to the azimuthal directions.
This drop is caused by the canceling out of these two components, whose polarization directions are orthogonal to each other.
In contrast, the polarization fraction along the major axis at a given orbital radius has a peak because the polarization components are in the same direction.
Outside of the peaks regions, the polarization fraction decreases with increasing radius because the radiation field becomes more isotropic along the major axis.

The other interesting feature is that the polarization vectors do not change when the dust components are changed but the polarization fraction does.
This is because the flux anisotropies determine the vector pattern, but the grain size does not change the flux anisotropies.
We can thus conclude that the grain size only affects the polarization fraction; it does not affect the polarization vectors in self-scattering.

In summary, the azimuthal variation is mainly determined by dust scale height and the absolute value of the polarization fraction can be tuned by changing the ratio of the two populations.
Based on these qualitative findings, we discuss the detailed modeling of HD 163296 in the next section.

\section{Application to HD 163296} \label{sec:target}

In the previous section, we showed that the polarization fraction of the azimuthal component increases with increasing dust scale height, enhancing the azimuthal variation.
The polarization observations of the HD 163296 disk shows such azimuthal variation, especially beyond the 70 au ring, which may indicate that dust grains do not fully settle to the midplane of the disk.
 Note that the azimuthal component of the polarization can be recognized by the Stokes {\it Q} image in Figure \ref{obs} because the Stokes {\it Q} emission represents polarization pointing along the disk major axis in this disk geometry. Therefore, the detection of the Stokes {\it Q} emission outside the 70 au ring indicates the polarization of the azimuthal component.
Here, we describe possible distributions of dust grains and dust scale heights in HD 163296.
First, we show the case for a single size distribution, then we consider two dust populations, and finally we discuss the dust scale height.

\subsection{Single Grain Population}

As a simple case, we consider a grain size population with a single power-law distribution with $n(a)\propto a^{-3.5}$ and $a_{\rm min}=0.1$ $\mu$m.
The maximum grain size $a_{\rm max}$ is the main parameter. We fix $a_{\rm max}=140$ $\mu$m, which produces polarization with the maximum efficiency.
We vary the dust scale height of the disk and investigate whether our model matches the observations.

\begin{figure}[htbp]
  \begin{center}
  \includegraphics[width=8.5cm,bb=0 0 2306 1175]{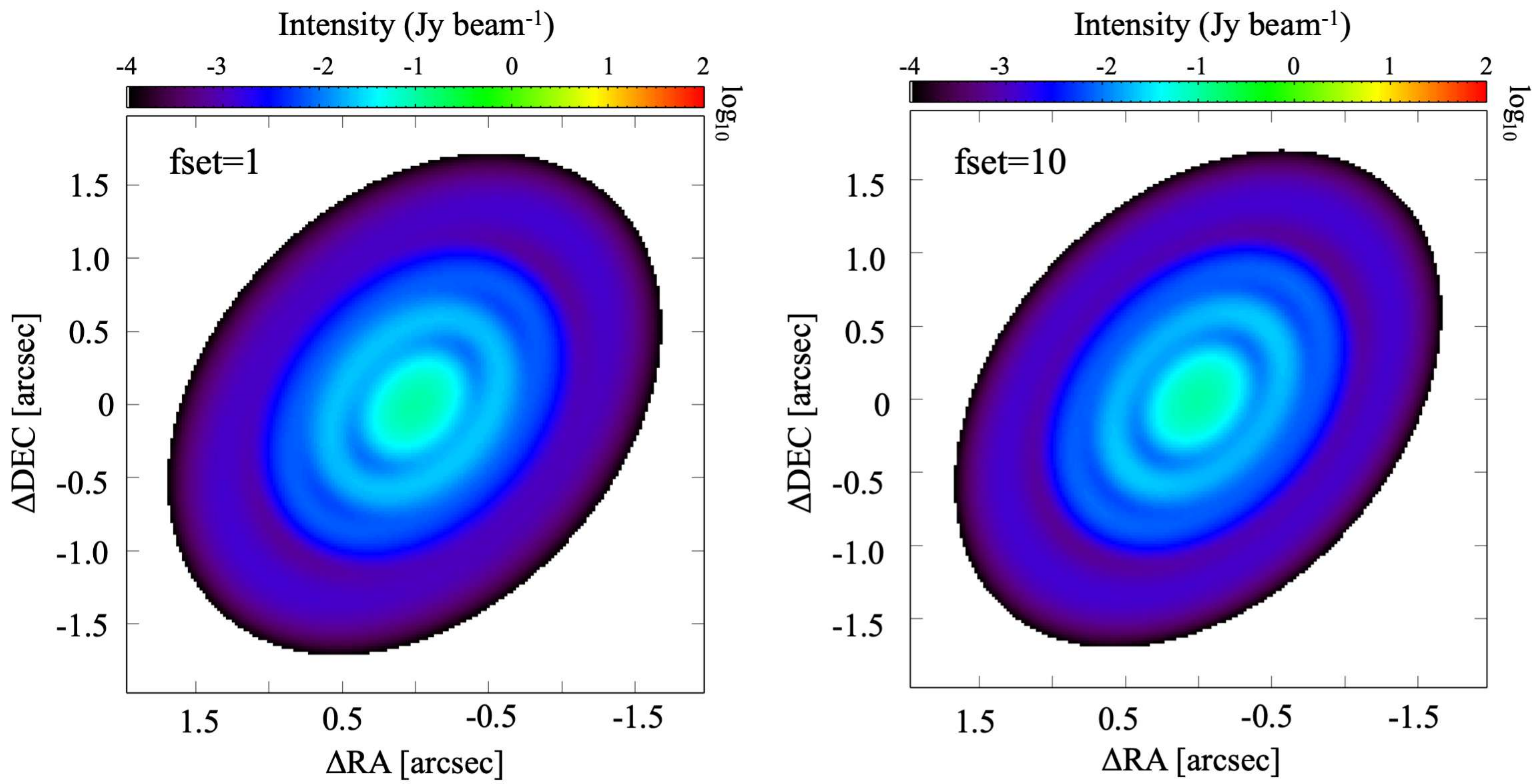}
  \end{center}
  \caption{Total intensity maps of radiative transfer calculations of HD 163296 with $f_{\rm set}=1$ and $10$.
  }
  \label{intensity}
\end{figure}

\begin{figure*}[htbp]
  \begin{center}
  \includegraphics[width=18cm,bb=0 0 2302 1676]{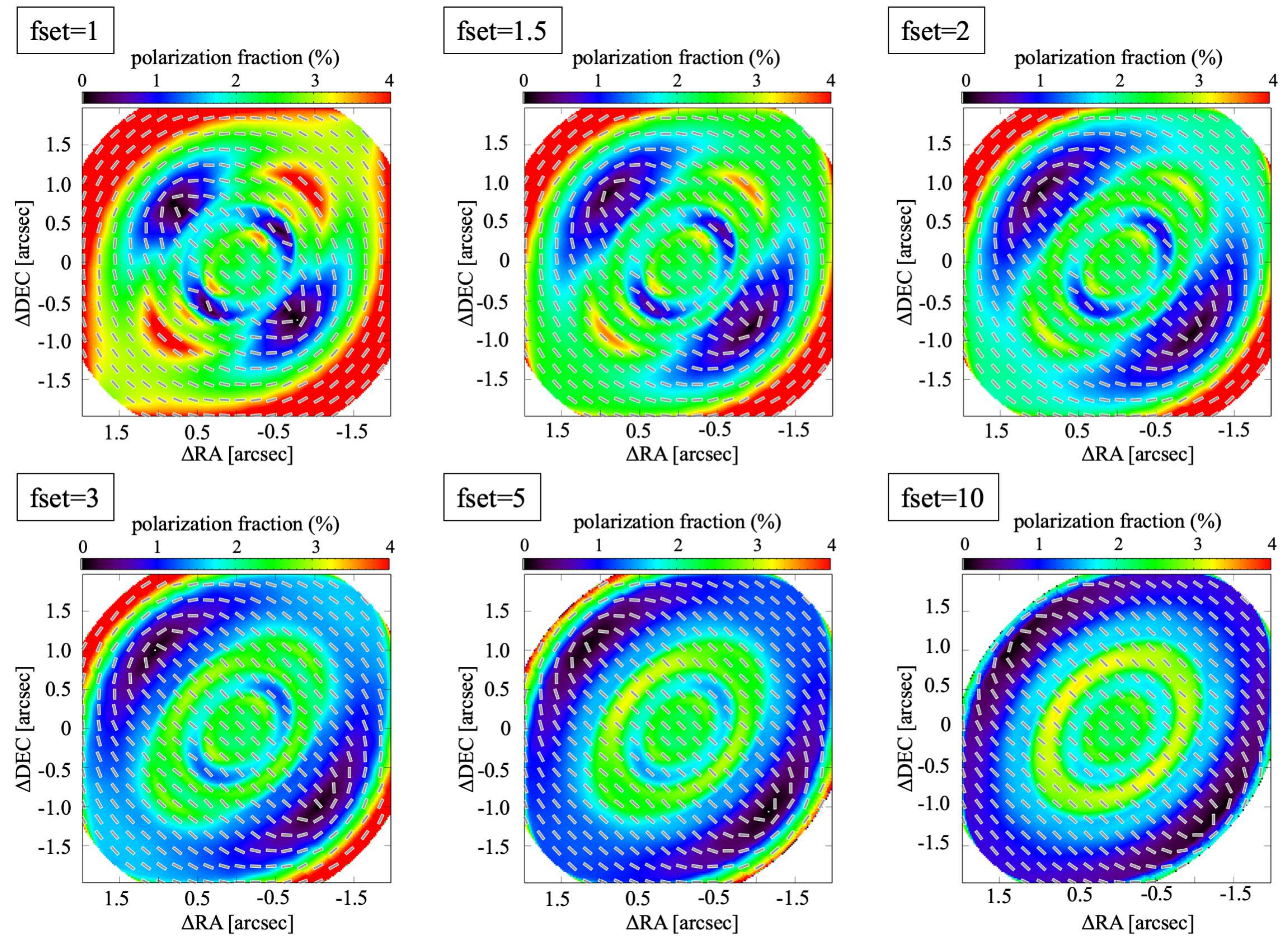}
  \end{center}
  \caption{Models of the polarization fraction and polarization vectors of HD 163296 for various dust scale heights. A single grain population with a size of 140 $\mu$m is distributed.
  }
  \label{fset}
\end{figure*}

First, we investigate the effects of dust scale height on the total intensity image.
Figure \ref{intensity} shows the total intensity maps with $f_{\rm set}=1$ and $10$.
The total intensity images for $f_{\rm set}=1$ and $10$ are almost identical and the projected ring width is different between the major and minor axes.
This is because the convolved beam size (0.2 arcsec) is insufficiently small to observe differences in the total intensity image.

\begin{figure}[htbp]
  \begin{center}
  \includegraphics[width=8.5cm,bb=0 0 2050 1423]{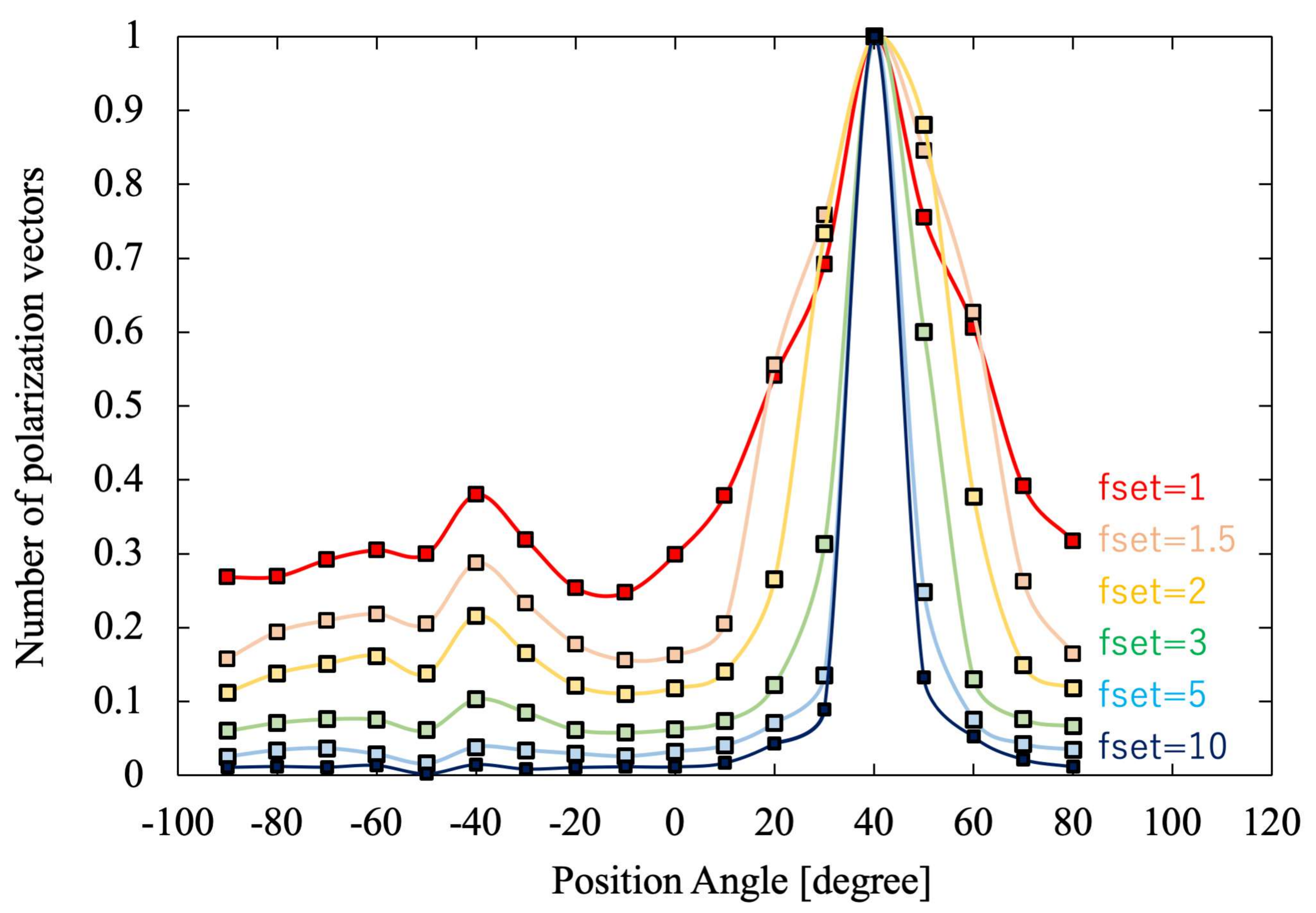}
  \end{center}
  \caption{Distribution of the polarization vectors in Figure \ref{fset} for models with various $f_{\rm set}$ values. The number of polarization vectors is normalized so that the maximum number is one.
  }
  \label{pdf}
\end{figure}

Second, we investigate the effects of dust scale height on the scattering-induced polarization.
Figure \ref{fset} shows the polarization fraction overlaid with the polarization vectors for various dust scale heights. 
In all cases, the polarization fraction is higher at locations in the gaps and lower at locations in the rings.
In addition, the polarization vectors are parallel to the disk minor axis in the center and in the azimuthal directions in the outer regions.
Different polarization patterns can be clearly seen, unlike the case for the total intensity.
The polarization vectors in all cases are the combination of those parallel to the minor axis in the center and those in the azimuthal directions in the outer regions. 
However, the contribution of the azimuthal pattern to the polarization decreases as the dust scale height decreases.
In addition, the variation of the polarization fraction in the azimuth direction decreases with decreasing dust scale height.
 To confirm the tendency of the azimuthal patterns of the polarization vectors to become enhanced with increasing dust scale height, we plot histograms of the polarization vectors for each model in Figure \ref{pdf}. The number of polarization vectors is normalized so that the maximum number is one. As shown in the figure, the polarization vectors are more widely distributed in position angle with increasing dust scale height. This is consistent with the above results.

Here, we discuss why the polarization fraction is high in the gaps and low in the rings.
To investigate the isotropicity of the radiation in the rings and gaps, we derive the physical length ($l$) where the scattering optical depth becomes unity.
The scattering optical depth is defined as
\begin{equation}
    \tau_{\rm scat}\equiv\kappa_{\rm scat}\rho_{\rm dust} l\sim \kappa_{\rm scat}\frac{\Sigma_{\rm dust}}{H_{\rm dust}} l,
\end{equation}
where $\kappa_{\rm scat}$ is the scattering opacity, $\rho_{\rm dust}$ is the spatial dust density, and $\Sigma_{\rm dust}$ is the dust surface density.
From this equation, we can estimate the physical length where $\tau_{\rm scat}=1$ as
\begin{equation}
    l\sim\frac{H_{\rm dust}}{\kappa_{\rm scat}\Sigma_{\rm dust}}\sim\frac{1}{f_{\rm set}}\frac{H_{\rm gas}}{ (\kappa_{\rm scat}/\kappa_{\rm abs})\tau_{\rm abs}},
\end{equation}
where $\tau_{abs}=\Sigma_{\rm dust}\times\kappa_{abs}$.
Therefore, $l$ decreases with $f_{\rm set}$  assuming that the surface density is fixed. The scattering is contributed to only by nearby radiation if dust grains are settled.

  We calculated the scattering and absorption opacities using Mie theory and derived to be $\sim4.5$ g cm$^{-2}$ and $\sim0.85$ g cm$^{-2}$, respectively. Here, we assumed the grain size of $a_{\rm max}=140$ $\mu$m and the observing wavelength of $\lambda=870$ $\mu$m.
At the 70 au ring of HD 163296, $l<H_{\rm gas}$ because the optical depth at the ring is $\tau_{\rm abs}\gtrsim 1$ and $\kappa_{scat}/\kappa_{abs}\sim 5$.
Therefore, the radiation field that contributes to scattering at the ring is almost isotropic because the width of the rings is similar to the gas scale height \citep{dul18}.
In contrast, the optical thickness at the gaps is $\tau_{\rm abs}\lesssim10^{-2}$.
Therefore, $l>H_{\rm gas}$, which means that the larger-scale radiation gradient contributes to the scattering and thus the radial radiation gradients produce radial anisotropy and azimuthal polarization.
This explains why the polarization fraction is higher in the gaps and lower in the rings.

\begin{figure}[htbp]
  \begin{center}
  \includegraphics[width=8.5cm,bb=0 0 830 1660]{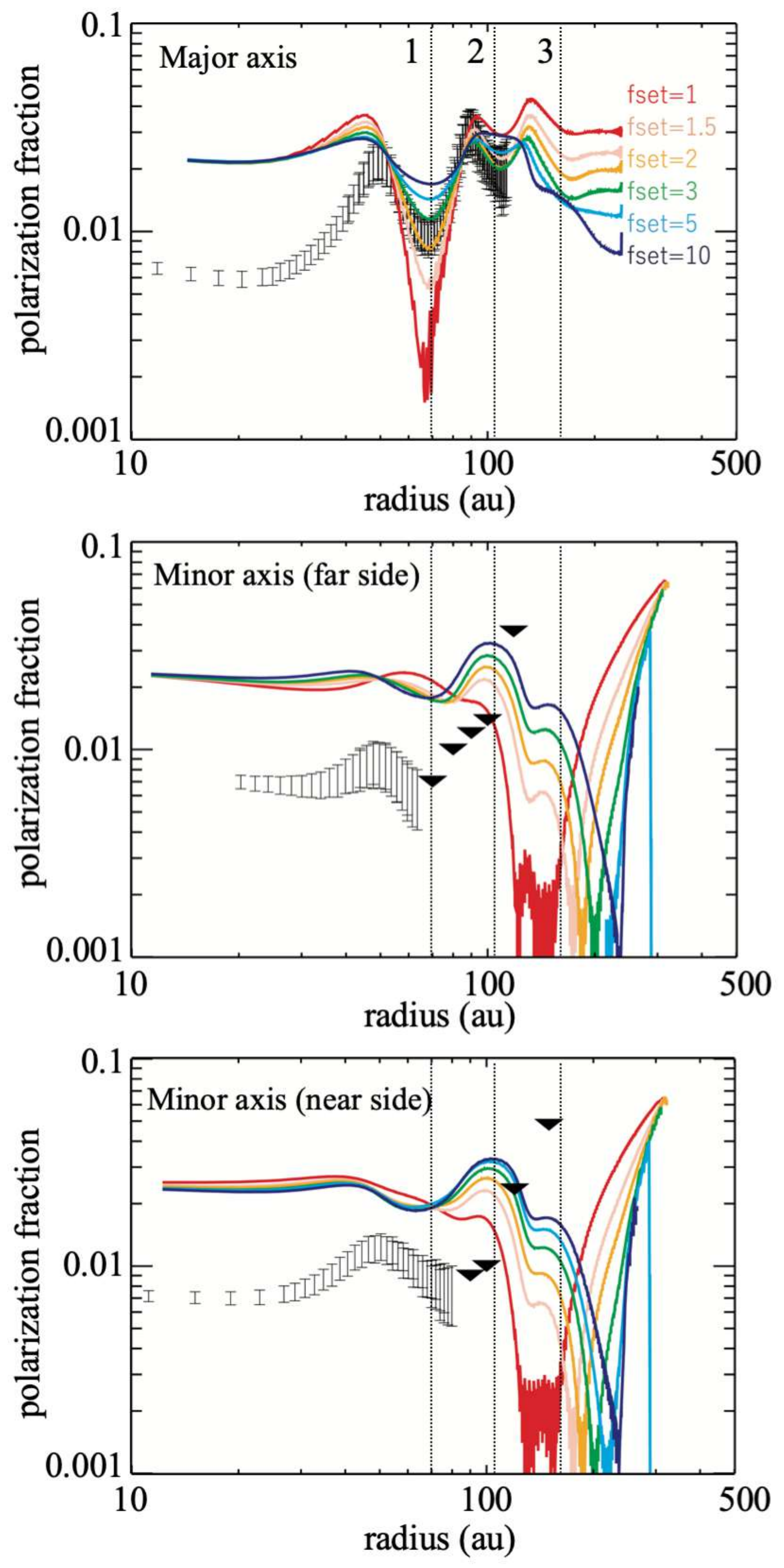}
    \end{center}
  \caption{Radial plots of the polarization fraction along the major and minor axes for models with various $f_{\rm set}$ values. Error bars indicate $\pm1\sigma$. Filled triangles represent the 3$\sigma$ upper limit.
  The models have a single grain population with a grain size of 140 $\mu$m.
  The radii of the three intensity rings are illustrated by the dotted vertical lines.
  }
  \label{radial_plot_140}
\end{figure}

\begin{figure*}[htbp]
  \begin{center}
  \includegraphics[width=17.cm,bb=0 0 2600 1687]{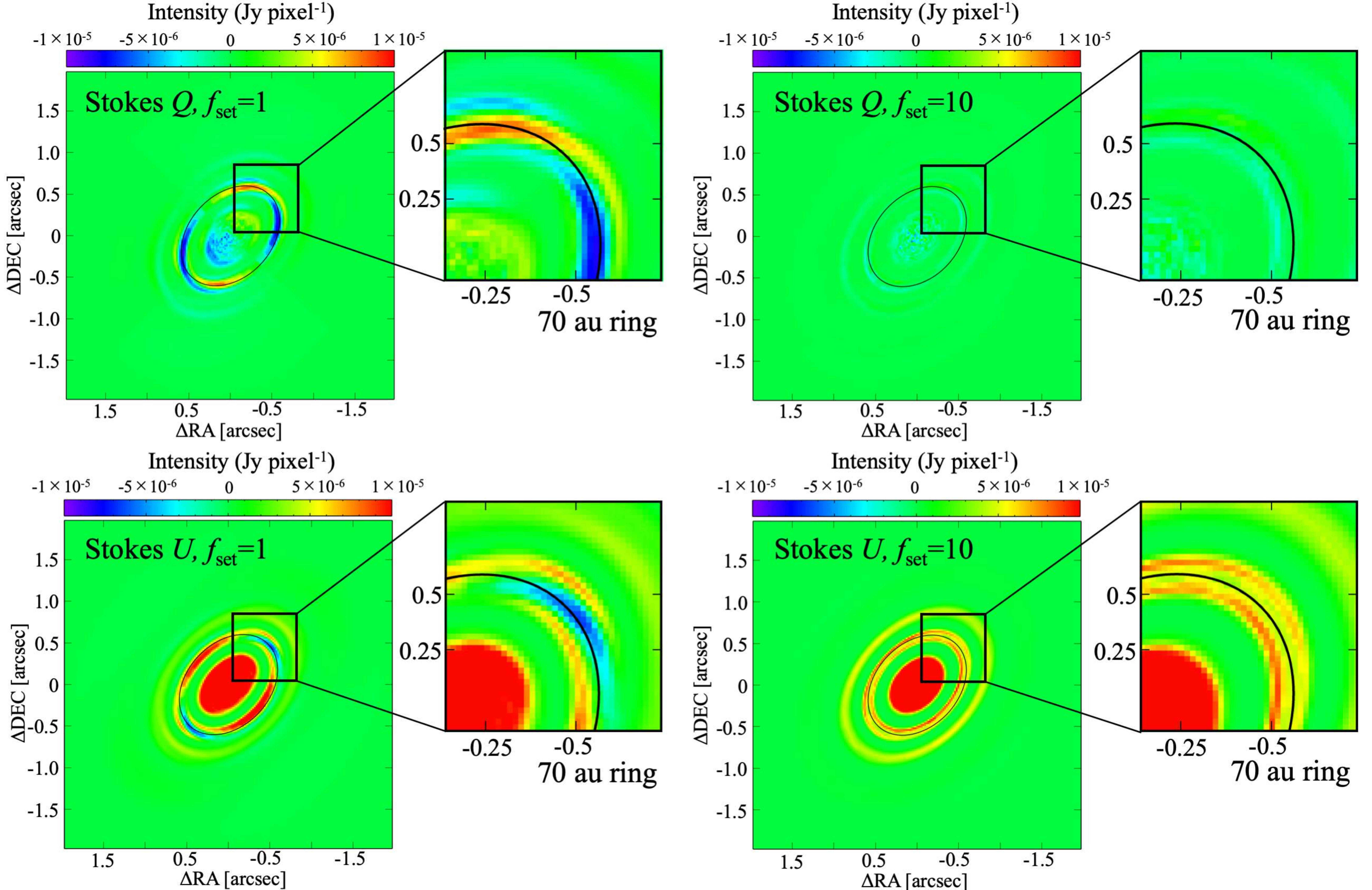}
  \end{center}
  \caption{ Synthetic images of the Stokes {\it Q} and {\it U} maps of HD 163296 with $f{\rm set}=1$ and $10$. These images are before convolution by the beam size. A single grain population with a grain size of 140 $\mu$m is present. The black circles show the 70 au ring. These images are not smoothed by the beam size.  The unit of the intensity is set to Jy pixel$^{-1}$. The pixel size is $0\farcs02$.
  }
  \label{QUmap}
\end{figure*}

Figure \ref{radial_plot_140} shows radial plots of the polarization fraction along the major and minor axes. The colors represent differences in dust scale height. The error bars indicate $\pm1\sigma$ of the observed polarization fraction.
As shown in the figure, along the major axis, the polarization fraction increases in the gaps and decreases in the rings with increasing dust scale height.
This opposite behavior is explained by canceling-out of the two orthogonal components of the scattering in the ring due to beam convolution.

In Figure \ref{QUmap}, to understand this complicated behavior, we show the {\it Q} and {\it U} maps for $f_{\rm set}=1$ and $f_{\rm set}=10$ before beam convolution. The black circle indicates the 70 au ring. 
In this set up, the Stokes {\it U} emission basically corresponds to the polarization vectors parallel to the minor axis and the butterfly-like pattern in the Stokes {\it Q} image corresponds to the azimuthal polarization vectors. 
The Stokes {\it Q} image for $f_{\rm set}=1$ shows emission at the 70 au ring, which indicates that the azimuthal component is produced. In contrast, the Stokes {\it Q} image for $f_{\rm set}=10$ shows almost no emission, which indicates that the polarization vectors are almost parallel to the disk minor axis.
The Stokes {\it U} image for $f_{\rm set}=1$ shows negative values only in the 70 au ring of the major axis; the other regions have positive values.
Therefore, for $f_{\rm set}=1$, the Stokes {\it U} intensity is decreased at the ring of the major axis due to beam dilution after beam convolution.
The polarization for $f_{\rm set}=1$ in Figure \ref{fset} shows that the polarization vectors have different directions on the scale of the beam size and that the polarization fraction drops in the ring of the major axis.

We now explain why the Stokes {\it U} image for $f_{\rm set}=1$ produces negative values in the ring.
The negative emission of the Stokes {\it U} in the ring of the major axis can be understood by regarding the ring as a thin disk, where the ring height is much smaller than the ring width, or a thick tube, where the ring height is comparable to the ring width.
For the thin disk, the scattering-induced polarization has polarization vectors parallel to the disk minor axis, as described in the previous section.
In contrast, as \citet{kat15} showed, the tube-like structure of the disk produces scattering-induced polarization pointing in radial directions on the ridge of the tube.
  This is because the net flux in the azimuthal direction is larger than that in the radial direction.

As shown in Figure \ref{radial_plot_140}, the polarization fraction along the minor axis at $r\gtrsim100$ au decreases with increasing dust scale height.
This is because of the canceling-out of the different orthogonal polarization directions (the azimuthal and disk minor axis directions), as described in Section 4.2.
Furthermore, the polarization fraction in the north-east direction is slightly higher than that in the south-west direction, which indicates that the polarization fraction of the near (northeast) side is higher than that of the far (southwest) side \citep{fla17}.
This is because the disk emission is marginally optically thick and the local inclination angles for the near and far sides are different \citep[see also Figure 6][]{yang17}.

\begin{figure}[htbp]
  \begin{center}
  \includegraphics[width=8.5cm,bb=0 0 830 1660]{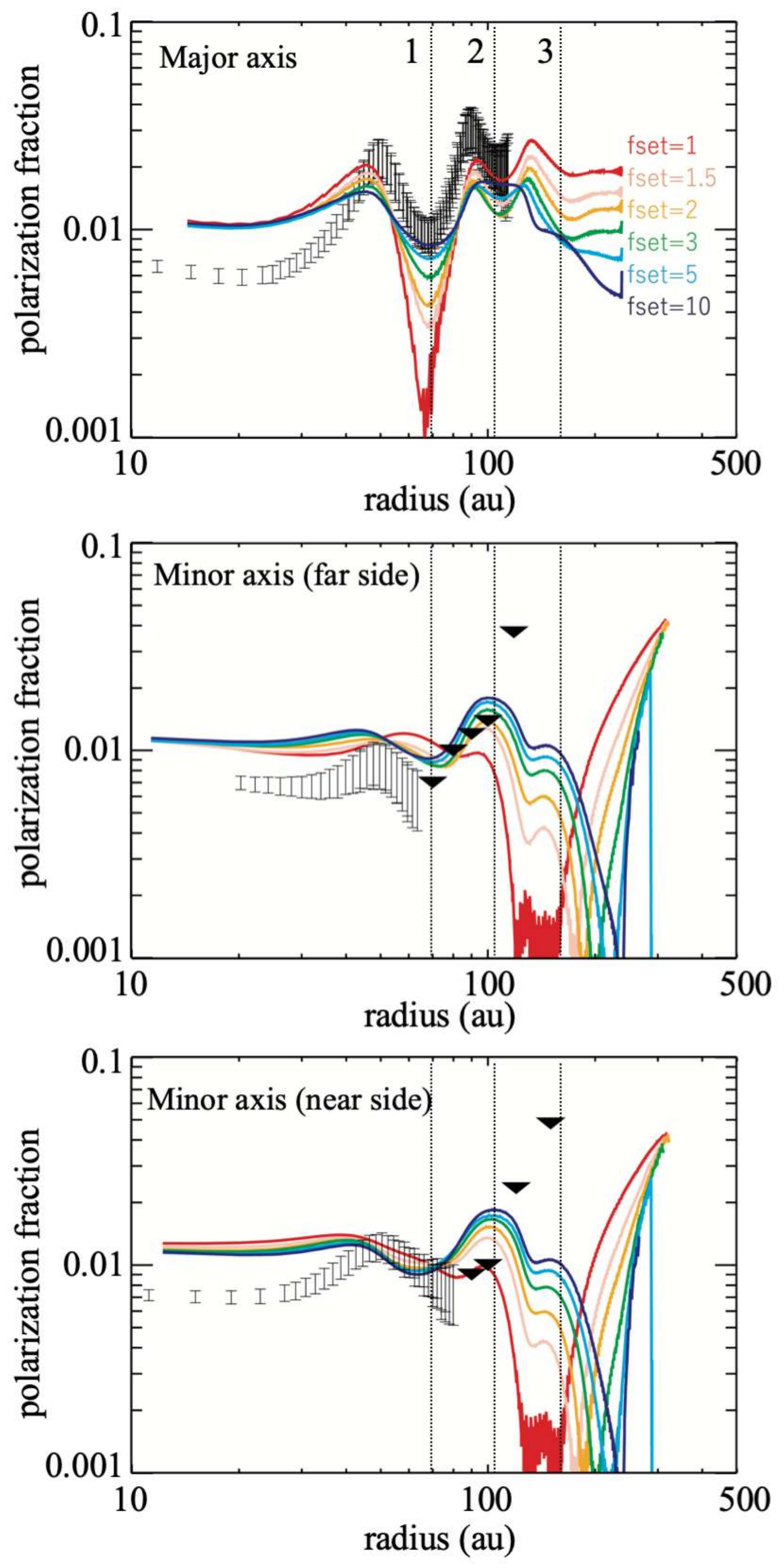}
    \end{center}
  \caption{Same as Figure \ref{radial_plot_140} but the models have a single grain population with a grain size of 160 $\mu$m.
  }
  \label{radial_plot_160}
\end{figure}

Here, we compare the observations and models.
As shown in Figure \ref{radial_plot_140}, along the major axis, the polarization fraction for $f_{\rm set}=2$ is well matched.
However, all $f_{\rm set}$ values do not match the observations along the minor axes for both near and far sides.
These models show a higher polarization fraction ($\sim2-3$\%) than the observation ($\sim0.5-1.5$\%) on the minor axis within $r\lesssim100$ au.
The low polarization fraction indicates that the grains are larger or smaller than 140 $\mu$m. Therefore, we performed the same calculations for a different grain size.

\begin{figure*}[htbp]
  \begin{center}
  \includegraphics[width=18cm,bb=0 0 2297 1676]{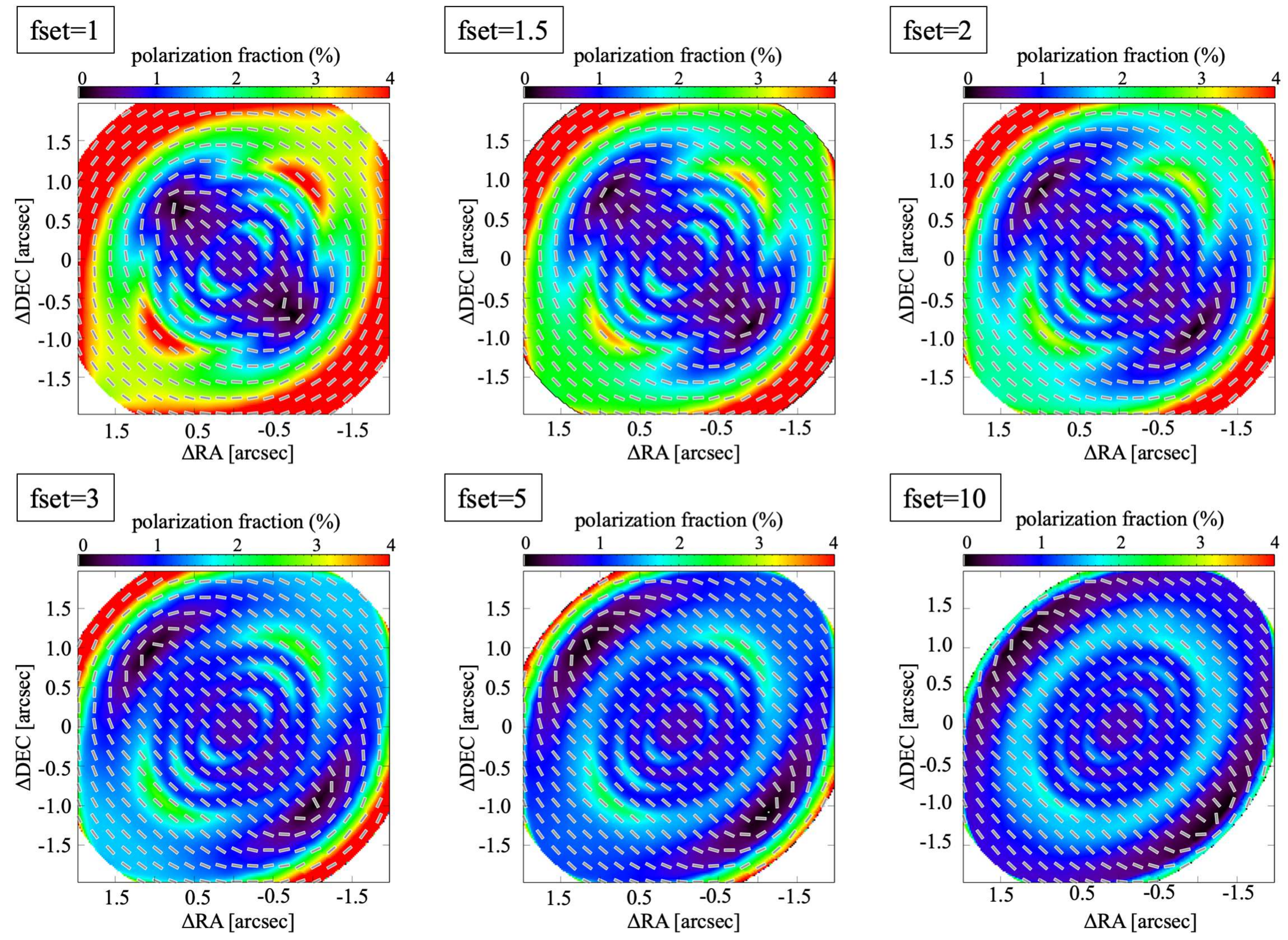}
  \end{center}
  \caption{Same as Figure \ref{fset} but the models have two populations of dust grains with grain sizes of 140 $\mu$m and 1 mm, respectively. The central part ($r\leq40$ au) is a combination of 140 $\mu$m and 1mm dust grains at a ratio of surface density (and density) of $5.5:4.5$, the rings with $\pm5$ au have 1 mm dust grains, and the gaps have 140 $\mu$m dust grains. The radii of the three intensity rings are illustrated by the dotted vertical lines.
  }
  \label{model1}
\end{figure*}

Figure \ref{radial_plot_160} shows the same plot as that in Figure \ref{radial_plot_140} but for a grain size of 160 $\mu$m for the polarization fraction to fit along the minor axis.
As shown in the previous section, the polarization vectors are the same as those in Figure \ref{radial_plot_140}, and the polarization fraction becomes lower when the grain size is changed. Even though the polarization fraction becomes lower than that in the previous models, the radial distributions of the polarization fraction do not match the observations for all $f_{\rm set}$. 
These models indicate that a model with a single grain population cannot reproduce the observations.

\subsection{Model with Two Dust Grain Populations}

Next, we investigate a model with two populations with different grain sizes.
Here, we assume that the two grain size populations are those shown in Figure \ref{grain_population}. The maximum grain sizes are set to 140 $\mu$m and 1 mm, respectively. The $140$ $\mu$m population contributes to both polarization and total intensity, whereas the 1 mm population contributes to only total intensity.

In the previous section, the observed polarization fraction in the gaps was well explained by the model with 140$\mu$m dust grains.
However, for the central part and the ring regions, there was inconsistency between the observations and models.
Therefore, we changed the dust grain populations in the central part and the rings while keeping the 140 $\mu$m population in the gaps.

We simply assume that the rings have only a 1 mm dust grain population because dust grains with a size on the order of 1 mm or larger have been suggested in studies of the spectral index $\beta$ \citep[e.g.,][]{ise07,ric10,tes14}.
We assume that the 1 mm dust grain population is distributed within $\pm5$ au from the centers of the rings.
This is consistent with ring sizes of 3.8 and 5.6 au derived from high resolution observations \citep{ise18}.

In the central part, which is within $r\leq40$ au, we use a mixture of the two dust grain populations (140 $\mu$m and 1mm). 
We set a ratio of surface density (and density) of $5.5:4.5$ (corresponding to a ratio of $11/9$) between the 140 $\mu$m and 1 mm dust grain populations to reproduce the observed polarization fraction.
Note that it is also possible to explain the lower polarization fraction by assuming a slightly larger ($\sim170$ $\mu$m) or smaller dust grain size ($\sim80$ $\mu$m) than 140 $\mu$m instead of the mixture of 140 $\mu$m and 1mm dust grains.

Figure \ref{model1} shows images of the polarization fraction overlaid with the polarization vectors of the models of the two populations of dust grains.
The panels show different dust scale heights.
These models show that the polarization fraction increases on the major axis and decreases on the minor axis with increasing dust scale height. 
This trend is the same as that in Figure \ref{fset} and can be explained by the combination or canceling out of the two components of the azimuthal and disk minor axis directions (see Section 4.2).

\begin{figure}[htbp]
  \begin{center}
  \includegraphics[width=8.5cm,bb=0 0 830 1660]{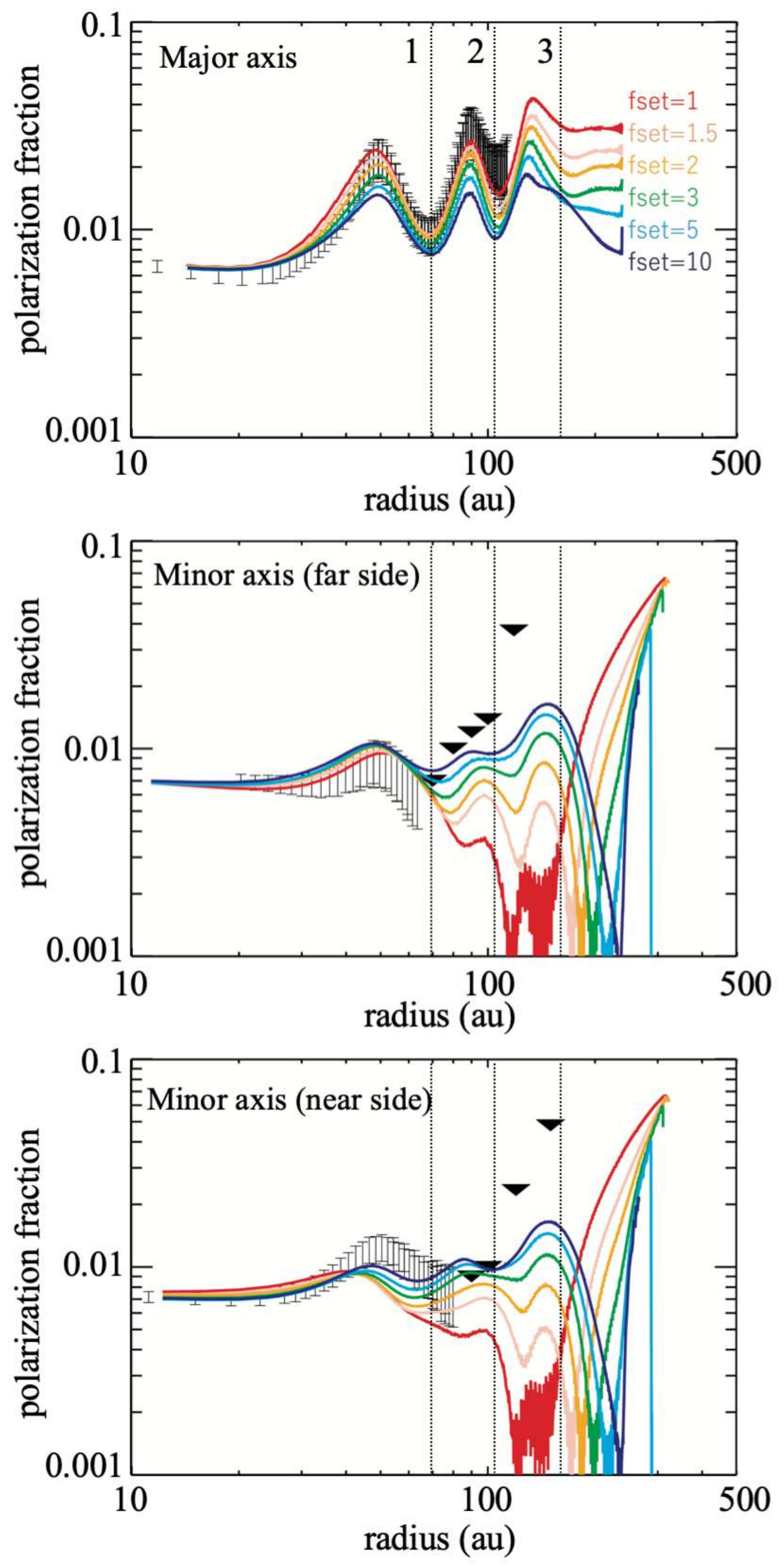}
    \end{center}
  \caption{Same as Figure \ref{radial_plot_140} but the models have two populations of dust grains, with grain sizes of 140 $\mu$m and 1 mm, respectively. The central part ($r\leq40$ au) is a combination of 140 $\mu$m and 1mm dust grains at a ratio of surface density (and density) of $5.5:4.5$, the rings with $\pm5$ au have 1 mm dust grains, and the gaps have 140 $\mu$m dust grains. The radii of the three intensity rings are illustrated by the dotted vertical lines.
  }
  \label{radial_plot_model1}
\end{figure}

Figure \ref{radial_plot_model1} shows radial plots of the polarization fraction along the major and minor axes, as done in Figures \ref{radial_plot_140} and \ref{radial_plot_160}.
A comparison of Figures \ref{radial_plot_140} and \ref{radial_plot_model1} indicates that the models with two populations of dust grains match the observations better than do the models with a single grain population.
For example, as shown in Figure \ref{radial_plot_140}, the polarization fraction along the major axis strongly depends on dust scale height for the single grain population.
In contrast, the models with two populations of dust grains are less sensitive to dust scale height and thus better reproduce the observations.
This is because the beam dilution does not affect the models with two populations of dust grains.
As discussed in the previous section, the drop is caused by the beam dilution of the different polarization directions on the scale of the beam size on the rings.
However, the 1 mm dust grains produce no polarized emission by scattering on the rings.

Although the models with $f_{\rm set}=1-2$ look similar to the observations in terms of the radial plots of the polarization fraction along the major and minor axes, these models do not exactly match the observations. 
For example, the radial profile of the minor axis of the near side shows that the polarization fraction of the models is slightly lower than the observations at $50\lesssim r \lesssim100$ au.

\begin{figure*}[htbp]
  \begin{center}
  \includegraphics[width=18.cm,bb=0 0 2999 1661]{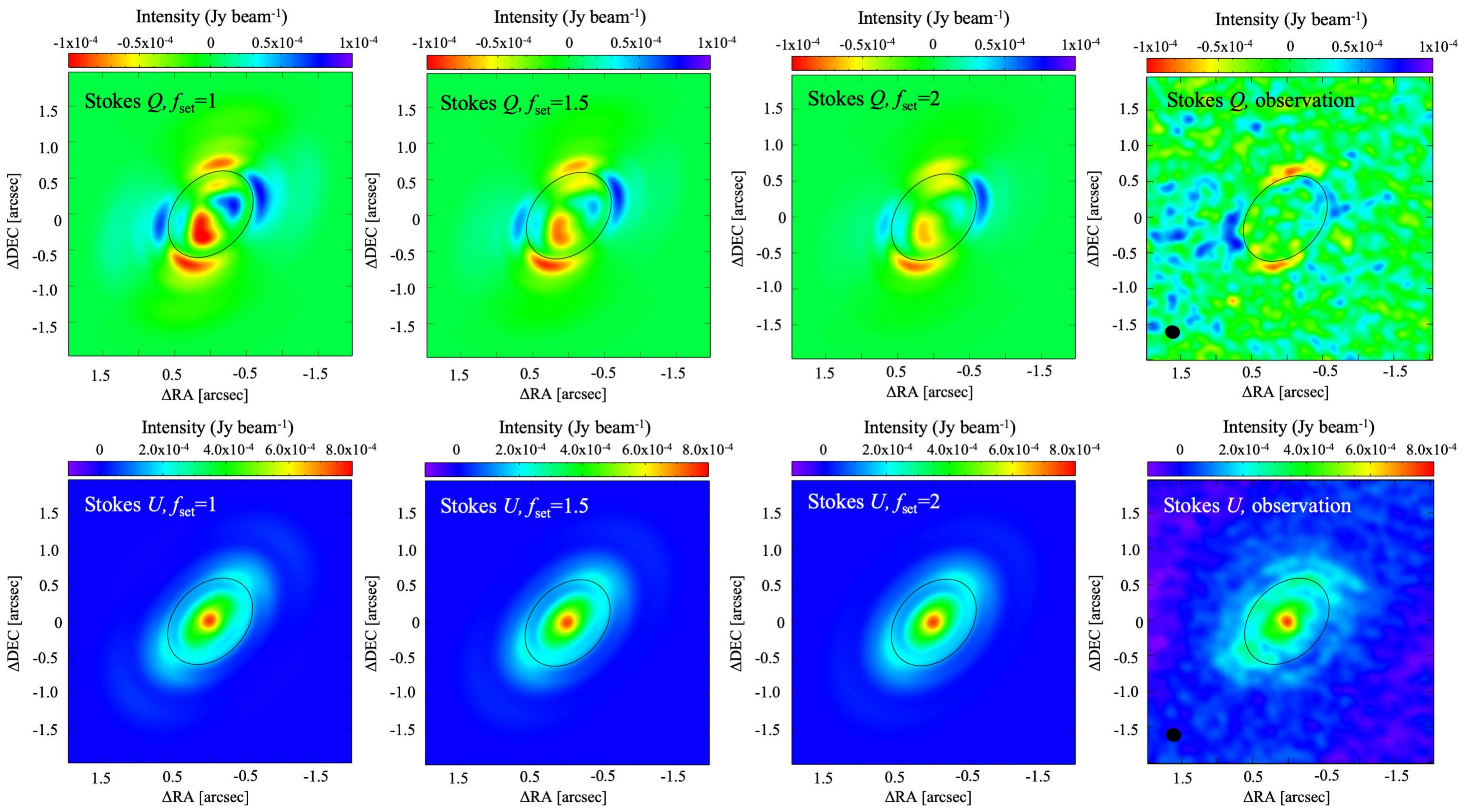}
  \end{center}
  \caption{Stokes {\it Q} and {\it U} images of the models with two populations of dust grains for various dust scale heights. The right panel shows the Stokes {\it Q} and {\it U} images of the observation. The black circles show the 70 au ring.
  }
  \label{QU_fset}
\end{figure*}

To compare the models directly with the observations, we investigate the Stokes {\it Q} and {\it U} images.
Differences in the Stokes {\it Q} and {\it U} images between the models with two grain populations with $f_{\rm set}=1-2$ and observations were found, as shown in Figure \ref{QU_fset}.
The observations show that the Stokes {\it Q} emission is only detected outside the 70 au ring with an intensity range of $-8.7\times10^{-5}$ Jy beam$^{-1}$ to $7.8\times10^{-5}$ Jy beam$^{-1}$.
In contrast, the models produce the Stokes {\it Q} emission inside the 70 au ring as well as outside the ring.
We investigated each model by comparing the absolute intensity value of the Stokes {\it Q} emission to fit the observations.
For $f_{\rm set}=1$, the Stokes {\it Q} emission ranges from $-1.1\times10^{-4}$ Jy beam$^{-1}$ to $7.8\times10^{-5}$ Jy beam$^{-1}$, which is a slightly wider intensity range than the observations both inside and outside the 70 au ring.
For $f_{\rm set}=1.5$, the Stokes {\it Q} emission ranges from $-8.5\times10^{-5}$ Jy beam$^{-1}$ to $7.6\times10^{-5}$ Jy beam$^{-1}$, which is similar to the observations outside the 70 au ring but is higher than the observations inside the ring. 
To reproduce the Stokes {\it Q} emission inside the ring, the dust scale height is needed to be further reduced. However, for $f_{\rm set}=2$, the Stokes {\it Q} emission ranges from $-8.1\times10^{-5}$ Jy beam$^{-1}$ to $6.7\times10^{-5}$ Jy beam$^{-1}$, which is a narrower intensity range than the observations outside the 70 au ring.
Therefore, we conclude that the radially constant dust settling parameter $f_{\rm set}$ cannot reproduce the observations.

 To decrease the Stokes {\it Q} emission, the maximum grain size can be made slightly larger (or smaller) than 140 $\mu$m or the population of the 1 mm dust grains can be increased.
However, these conditions would also decrease the Stokes {\it U}, resulting in lower polarization fraction. The non-detection of the Stokes {\it Q} requires dust grains larger than 200 $\mu$m or the mass ratio of 140 $\mu$m grains to 1 mm grains to be $0.4:0.6$ with assuming that $f_{\rm set}=1.5$. These dust grains give a polarization fraction of $\lesssim0.4$\% in the central part, which is lower than the observations.
Therefore, the only way to reduce the Stokes {\it Q} emission keeping the Stokes {\it U} emission is to change the dust scale height.
Therefore, the low Stokes {\it Q} emission cannot be explained by larger dust grains or a different mass ratio between the two grain sizes.

\subsection{Radial variation of dust scale height}
In the previous subsection, we showed that the models with two populations of dust grains match the observations better than do the models with a single grain population.
However, there are still inconsistencies between the models and observations in terms of the radial profile and the Stokes {\it Q} and {\it U} distributions.
Here, we introduce the radial variation of dust scale height to improve the models.

\begin{figure}[htbp]
  \begin{center}
  \includegraphics[width=8.5cm,bb=0 0 1856 1387]{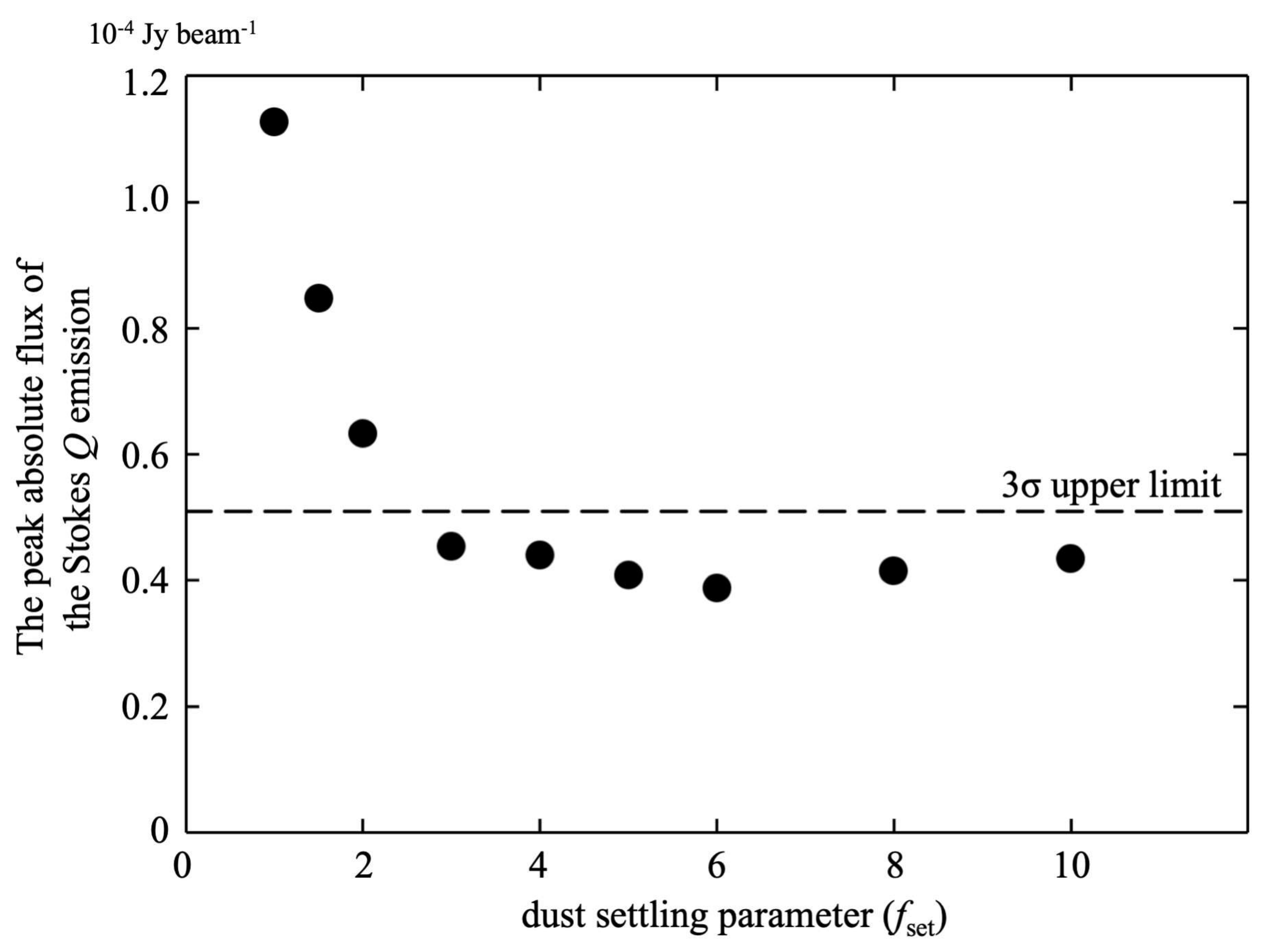}
    \end{center}
  \caption{Peak absolute flux of the Stokes {\it Q} emission inside the 70 au ring versus the dust settling parameter ($f_{\rm set}$). The intensity is smoothed by a beam size of 0.2 arcsec.
  }
  \label{fset_q}
\end{figure}

\begin{figure*}[htbp]
  \begin{center}
  \includegraphics[width=18.cm,bb=0 0 2378 1686]{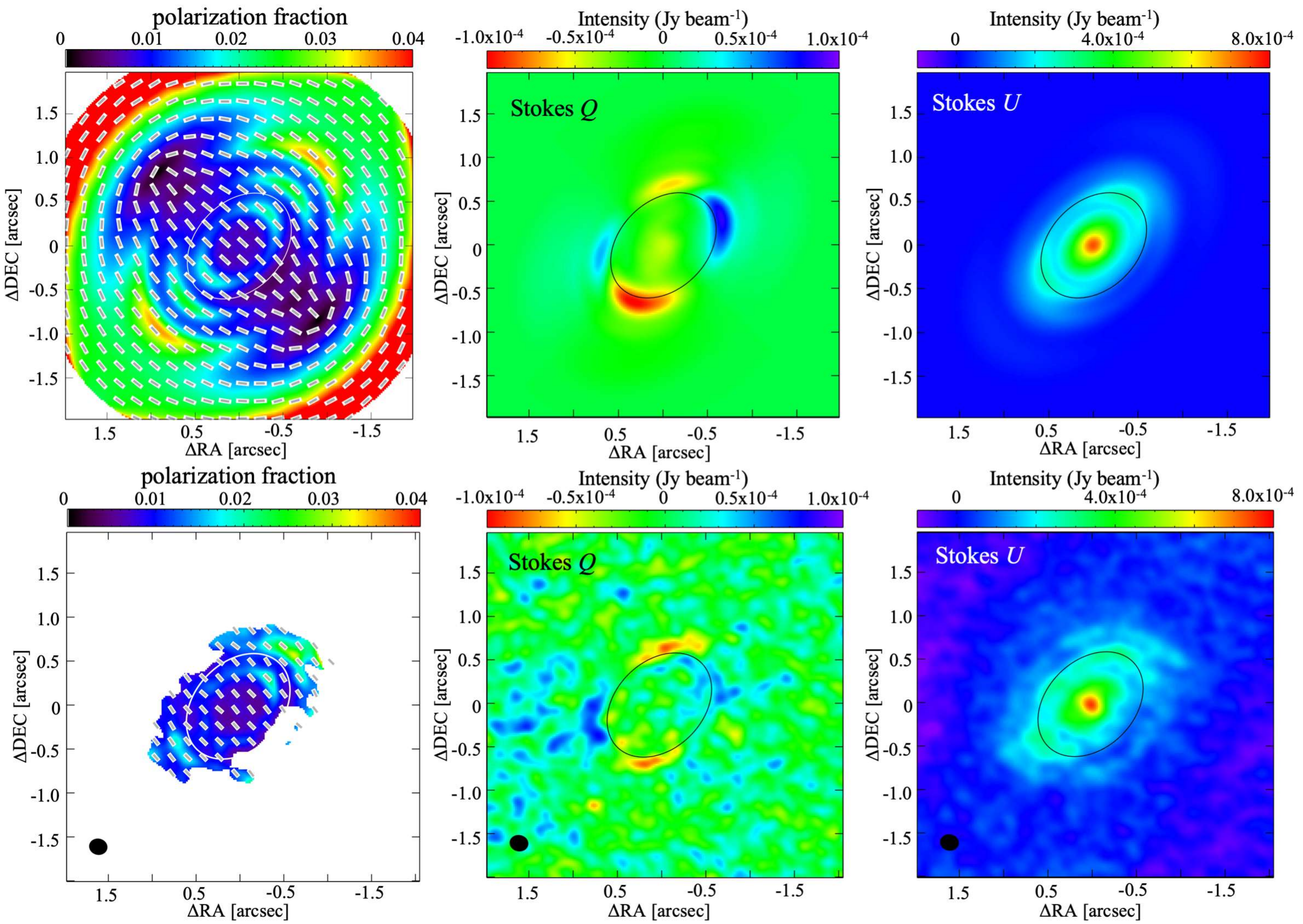}
  \end{center}
  \caption{Upper panels show model images of the polarization fraction and Stokes {\it Q} and {\it U}. The model has two populations of dust grains and the dust scale height ranges from ${\rm fset}=1.5$ to ${\rm fset=10}$ at 68 au. The lower panels show the polarization fraction and Stokes {\it Q} and {\it U} images from the observations. The white and black circles show the 70 au ring.
  }
  \label{comparison}
\end{figure*}

Outside the 70 au ring, the model with $f_{\rm set}=1.5$ is consistent with the observations. However, the model with $f_{\rm set}=1.5$ shows higher Stokes {\it Q} intensity than the observations inside the 70 au ring.
As shown before, the Stokes {\it Q} emission is reduced by decreasing the dust scale height. Therefore, we need to reduce the dust scale height inside the 70 au ring.
We investigate the dust settling parameter ($f_{\rm set}$) inside the 70 au ring to explain the non-detection of the Stokes {\it Q} emission.
Figure \ref{fset_q} shows the absolute value of the peak intensity of the Stokes {\it Q} emission inside the 70 au ring for each model with $f_{\rm set}=1-10$. The absolute value is used because the Stokes {\it Q} emission has both positive and negative values.
As shown in the figure, the Stokes {\it Q} emission is reduced by decreasing the dust scale height. The  Stokes {\it Q} emission may saturate at $f_{\rm set}\sim6$ because it is also produced by the component of the disk minor axis. The position angle of the disk is set to be 133.3 degrees. Therefore, the Stokes {\it Q} emission is slightly produced by the polarization of the disk minor axis even if the azimuthal component is not produced.
By taking into account the non-detection of the Stokes {\it Q} emission inside the 70 au ring, we can set the upper limit of the dust scale height as $H_{\rm dust}\lesssim (1/3)H_{\rm gas}$ from the 3$\sigma$ upper limit of the Stokes {\it Q} observations.

To reproduce the observations, we apply $f_{\rm set}=10$ within $r\leq68$ au and $f_{\rm set}=1.5$ for $r>68$ au.
Figure \ref{comparison} shows a comparison between our model and the observations. We compare the images not only for the polarization fraction but also for the Stokes {\it Q} and {\it U} distributions.
The model well matches the observations not only for the polarization fraction but also for the Stokes {\it Q} and {\it U} images.
Furthermore, the twist of the polarization vectors pointed out by \citet{den19} is reproduced.
Therefore, the polarization vectors in our model are consistent with the observations.
The original images of the Stokes {\it I}, {\it Q}, and {\it U} and the polarized intensity before convolution by the Gaussian beam are shown in Appendix \ref{original_image}.

\begin{figure}[htbp]
  \begin{center}
  \includegraphics[width=8.5cm,bb=0 0 830 1660]{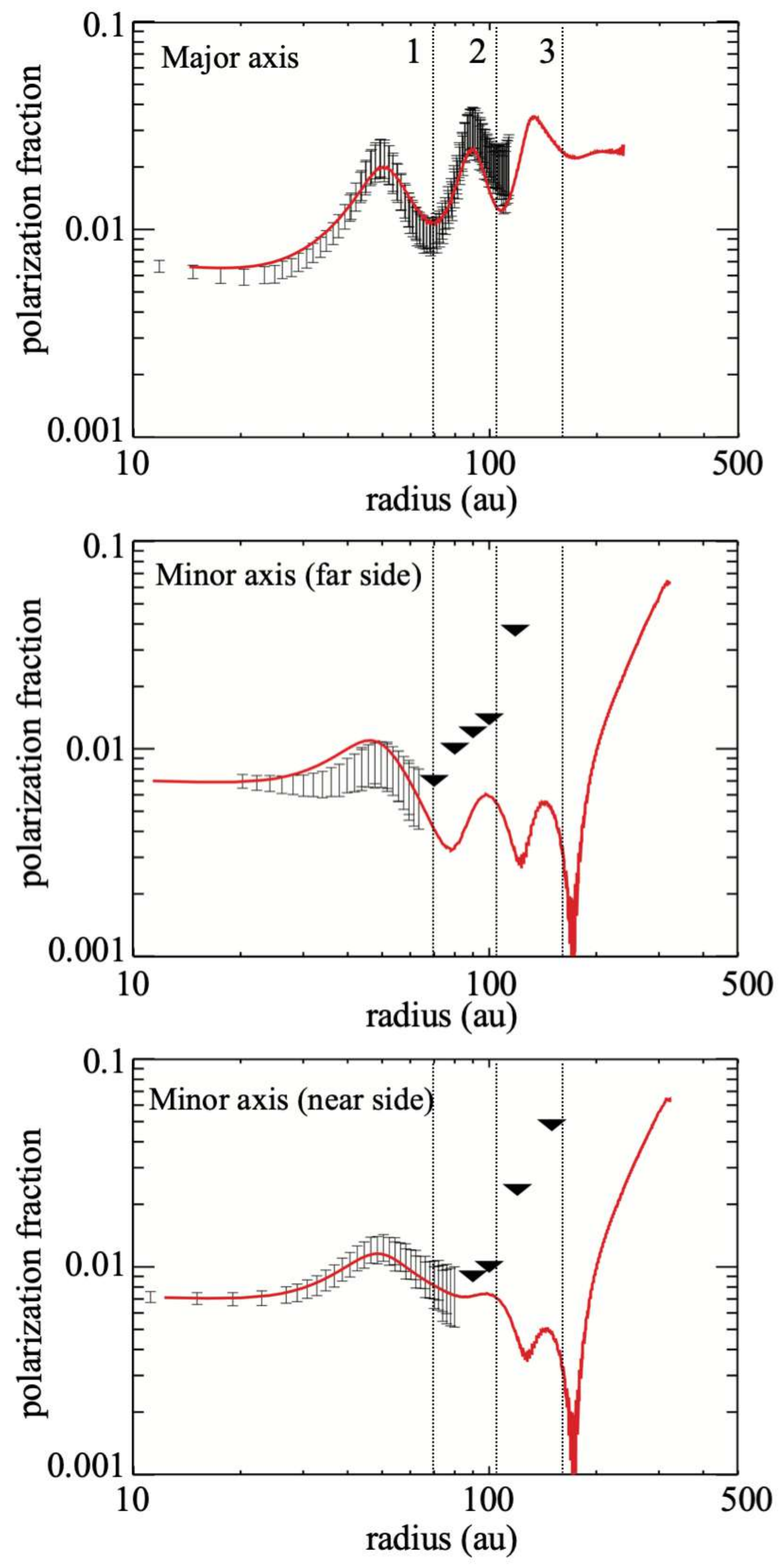}
    \end{center}
  \caption{Same as Figure \ref{radial_plot_140} but for our best model with two populations of dust grains. The dust scale height ranges from ${\rm fset}=1.5$ to ${\rm fset=10}$ at 68 au. The radii of the three intensity rings are illustrated by the dotted vertical lines.
  }
  \label{model_best_plot}
\end{figure}

We show radial plots of the polarization fraction for the best model and observations in Figure \ref{model_best_plot}.
The figure indicates that the model well matches the observations.
In particular, the radial profile along the minor axis of the near side is improved compared to that in Figure \ref{radial_plot_model1}.
This may be because the radiation becomes more anisotropic due to the gradient of the dust scale height.

\begin{figure}[htbp]
  \begin{center}
  \includegraphics[width=8.5cm,bb=0 0 1560 1220]{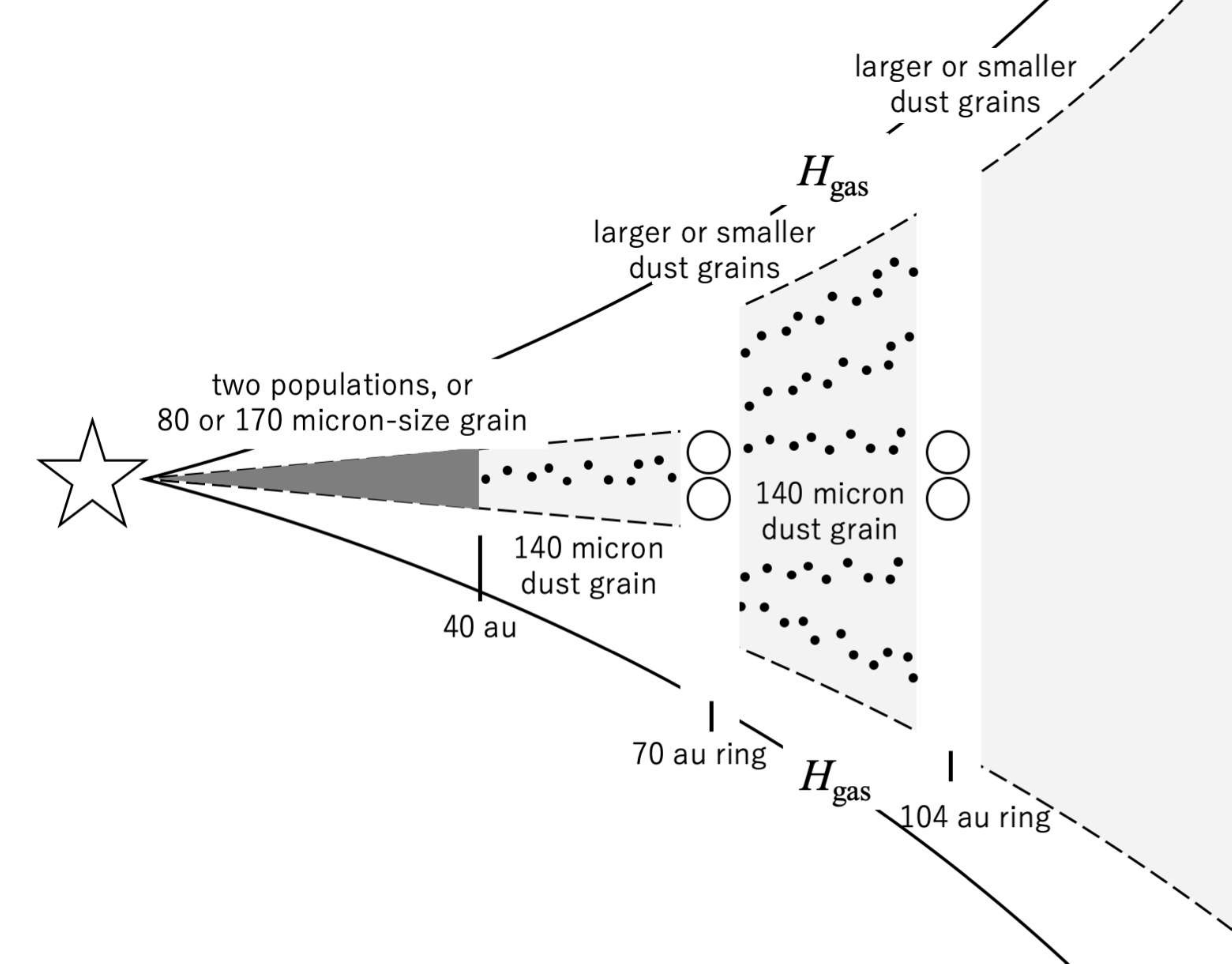}
    \end{center}
  \caption{Schematic view of distributions of dust grains and dust scale heights in the protoplanetary disk around HD 163296.
  }
  \label{summary}
\end{figure}

Figure \ref{summary} shows a schematic view of the grain size distribution and the dust scale height for our best model.
The best model is as follows. The 140 $\mu$m dust grains are distributed in the gaps, and dust grains contributing no polarization are distributed in the rings. In the rings, we simply assume 1 mm dust grains as a representative population that does not produce polarization at the observation wavelengths. However, 1 mm dust grains are just one possible example. The constraint from the polarization observations is the exclusion of grains with a size of around 140 $\mu$m. There are two possibilities for the central part because the observed polarization fraction is less than the prediction with 140 $\mu$m but still detected. One possibility is two populations of dust grains with grain sizes of 140 $\mu$m and 1 mm and a ratio of surface density of $5.5:4.5$. The other possibility is a grain size distribution that can be expressed as a single power law; the maximum grain size would be $\sim170$ $\mu$m or $\sim80$ $\mu$m rather than 140 $\mu$m.
Furthermore, the dust scale height inside the 70 au ring is at least three times smaller than the gas scale height, but it is two-thirds the gas scale height outside the 70 au ring. 
Note that we cannot constrain the dust scale height in the rings because the polarization due to self-scattering is not produced in our model.
An additional test of the single grain population with changing the dust scale height is shown in appendix \ref{ad_test}.
As shown in the appendix \ref{ad_test}, the observation results cannot be reproduced if the radially constant particle size is assumed even if the dust scale height is changed.

\section{Discussion}\label{sec:discussion}

In this section, we discuss the implication of these constraints and their physical origin.

\subsection{Distributions of dust grains}

We discuss the consistency in grain sizes derived from polarization and spectral index analyses. 
\citet{den19} derived the spectral index, $\alpha$, using the ALMA 0.87 and 1.3 mm dust continuum observations and showed that it is significantly less than $2.0$ in the central region, $\alpha\sim2.0$ in the 70 and 104 au rings, and slightly higher in the gaps than in the rings.
In general, a low spectral index can be explained by optically thick emission or large dust grains.
However, a spectral index of lower than 2 can be explained by the effect of dust scattering.
The continuum thermal dust emission becomes fainter than the black-body radiation if the scattering is efficient \citep[e.g.,][]{liu19,zhu19}.
This high efficiency of dust scattering at a 0.87 mm wavelength is consistent with the detection of self-scattering polarization. Therefore, a consistent grain size distribution between the spectral index and the polarization is as follows.
The central region inside 40 au, where self-scattering is efficient, but not maximally efficient, and the spectral index is lower than 2, is likely to be dominated by $80-170$ $\mu$m dust grains.
The rings, which have a low spectral index and polarization consistent with 1 mm grains, can be explained in two ways. (i) The rings are optically thick and the grain size is much larger or smaller than 140 $\mu$m; (ii) the rings are optically thin and the grain size is much larger than  140 $\mu$m.
The case with the larger grains distributed in the rings is consistent with the dust trapping scenario \citep[e.g.,][]{dul18}. In contrast, the case with the smaller grains distributed in the rings is consistent with the grain growth scenario of the nonsticky ice such as CO$_2$ \citep{oku19}.
The gaps, where the spectral index is high ($\alpha\sim2.5$) and self-scattering is very efficient, are dominated by 140 $\mu$m grains.

Although the radial grain size distribution with large grains in rings and small grains in gaps is generally consistent with the radial dust trapping scenario, it is still not fully consistent with grain growth and migration theory.
The synthetic observations of dust trapping and a grain growth/migration simulation with a Jupiter-mass planet show that the polarization fraction is only $\lesssim1$\% at the peaks around the gaps \citep{poh16}.
This is because the grain size distribution is not radially constant in gaps.
In contrast, the 0.87 mm polarization of HD 163296 shows that the peaks polarization fraction is $\sim4$\% in the gap center. 
The polarization results are more consistent with 140 $\mu$m dust grains widely distributed in the gaps.
For example, shallow bumps in the gas for a less massive planet would produce a shallow grain size distribution.

\subsection{Estimate of turbulence strength}
The dust scale height is determined by the balance between dust settling and gas turbulence.
In previous sections, we constrained the dust scale height and dust grain size based on the polarization measurements. Here, we constrain the gas turbulence.
The dust scale height is constrained to be $H_{\rm dust}/H_{\rm gas}\lesssim1/3$ for the $r\lesssim70$ au region, and $H_{\rm dust}/H_{\rm gas}\sim2/3$ for the $r\gtrsim 70$ au region.
By assuming a balance between vertical settling and turbulent diffusion, the dust scale height is written as \citep[e.g.,][]{dub95,you07}
\begin{equation}
        H_{\rm dust}=\Big(1+\frac{\rm St}{\alpha}\frac{1+2 \rm St}{1+\rm St}\Big)^{-1/2} H_{\rm gas},
\label{eq_scaleheight}
\end{equation}
where $H_{\rm gas}$ is the gas scale height $H_{\rm gas}=c_s/\Omega$ and St is the Stokes number. ${\rm St}=(\pi\rho a_{\rm dust})/(2\Sigma_{\rm gas})$. 
This equation yields $H_{\rm dust} \simeq\sqrt{{\alpha}/{\rm St}} H_{\rm gas}$ if we assume that ${\rm St}$ is much less than unity.
Using equation (\ref{eq_scaleheight}), we derive ${\rm St}/\alpha \gtrsim 8$ inside the 70 au ring and ${\rm St}/\alpha \simeq 1.25$ outside the 70 au ring; these values differ by about an order of magnitude.

To find the radial difference in turbulence, we discuss the turbulence strength in the two gaps at 50 and 90 au.
We conclude that these gaps are filled with grains with a size of 140 $\mu$m.
The gas surface density has been constrained by multiple CO line observations to be $3-10 {\rm~g~cm^{-2}}$ in the 50 au gap and $0.1-2 {\rm~g~cm^{-2}}$ in the 90 au gap \citep{ise16}.
Using these values and assuming that the dust material density is $\rho\sim1.67$ g cm$^{-3}$, the turbulence strength $\alpha$ is derived to be $\alpha\lesssim 1.5\times10^{-3}$ in the 50 au gap and $\alpha\simeq0.015-0.3$ in the 90 au gap.

The low value of $\alpha\lesssim1.5\times10^{-3}$ inside the 70 au ring is consistent with the upper limit derived from molecular line observations where $\alpha < 10^{-3}$ \citep{fla15,fla17}. 
Magnetohydrodynamic (MHD) simulations have shown that there is a floor value of turbulence strength of $\alpha \gtrsim 10^{-4}$ as long as the surface layers are magneto rotational instability (MRI)-active \citep{suz10,sim15}.
Even without magnetic fields, there are hydro instabilities that make the disk turbulent, such as vertical shear instability, which has a turbulence strength of $\alpha \sim 10^{-4}$ \citep{nel13}. \citet{lin19} study the dust settling against the vertical instability by taking into account the dust back-reaction onto the gas. They show that the dust can settle down to $0.1H_{\rm gas}$.
Therefore, the upper limit of $\alpha\lesssim1.5\times10^{-3}$ is consistent with the simulations of low-turbulence disks.

The turbulence strength of $\alpha\simeq0.015-0.3$ outside the 70 au ring corresponds to $v_{\rm turb}/c_s\simeq0.12-0.55$ by assuming $\alpha\sim(v_{\rm turb}/c_s)^2$, which is higher than the upper limit of $v_{\rm turb}/c_s\lesssim0.04-0.05$ toward the midplane derived from the molecular line observations \citep{fla17}.
One possible reason for these different estimations of turbulence strength is different observation regions between the molecular line and polarization observations.
The molecular line observations mainly trace the ring structures where the dust and gas surface densities are much higher than those in the gaps. The intensities of continuum and molecular lines are also high. Therefore, the line profile obtained with a larger beam size is dominated by the emission from the ring structures.
In contrast, the polarization observations allow us to derive the turbulence strength only in the gaps because the dust grains in the gaps dominate the polarization.
Therefore, the turbulence strength estimated from the molecular line and polarization observations may be for radially different regions of the disk.
Even if the dust settling parameter ($f_{\rm set}$) is the same for the rings and gaps, the turbulence strength $\alpha$ in these regions could be different because the gas surface densities are different.

It should be noted that the gas surface density was estimated from the CO observations with the assumption of a cosmic molecular abundance ($[{\rm ^{12}CO}]/[{\rm H_2}]=5\times10^{-5}$). If the CO abundance is lower or the CO intensity is reduced but the gas surface density does not drop in the dust gaps, the gas surface density will be higher than that used here. Then, the Stokes number and turbulence $\alpha$ will be lower.
 Actually, \citet{boo19} detected the rarest stable CO isotopologue, $^{13}$C$^{17}$O, in the disk and derived the gas mass with a factor of $2-6$ larger than previous gas mass estimates using C$^{18}$O.
In this case, the turbulence strength may be decreased up to be $\alpha\lesssim2.5\times10^{-4}$ in the 50 au gap and $\alpha\sim0.002$ in the 90 au gap, respectively.

Yet, the difference of an order of magnitude in turbulence strength between the inner and outer regions is robust.
One of the consistent scenarios for low turbulence in the inner region and high turbulence in the outer region is a dead zone of MRI \citep{gam96}. The dead zone is considered as a region where the MRI is stabilized by ohmic dissipation \citep{san99}. 
Furthermore, the MRI is also suppressed by the other non-ideal MHD effects of the ambipolar diffusion \citep{des04,bai11,dzy13} and the Hall effect \citep{war99,bal01,bai14}.
Theoretical calculations suggest that the outer edge of the dead zone  is located around $10-20$ au \citep{san00,sem04,bai09,dzy13}.
In contrast, the polarization observations suggest that the boundary between turbulence active and inactive zones is located at the 70 au ring, which is much larger than 20 au.

The radial transition from low to high turbulence strength may be explained by the electron heating (e-heating) zone rather than the dead zone \citep{oku15}.
In the e-heating zone, electrical conductivity is suppressed by heating of electrons due to turbulence.
As a result, the turbulence strength is significantly reduced.
Therefore, the e-heating zone may explain the low turbulence in the region where MRI is expected to be active.
\citet{mor16} showed that the e-heating zone can extend to $\sim100$ au and that the dead zone can reach $<50$ au depending on the magnetic field, grain size, and gas density. Therefore, the boundary between the turbulence-active and -inactive zones at the 70 au ring may indicate the transition from the e-heating zone to the active zone.

Note that in terms of the ring formation process, the dead zone scenario is qualitatively consistent with dust accumulation at the outer edge of the dead zone, where dust grains are accumulated at the gas pressure bump created by the dead zone outer edge \citep{flo15,pin16}.

\subsection{Comparison with SPHERE and HST observations}

\citet{mur18} observed the disk surface of HD 163296 in scattered light at {\it H} and {\it J} bands with the SPHERE instrument.
Based on the lack of scattered light emission detected in the rings at 104 and 160 au, they proposed two models. 
In the first model, small dust grains (a few $\mu$m) settle into the midplane in the outer disk and thus the outer regions are in the shadow cast by the 70 au ring.
In the second model, there is a depletion of dust grains smaller than $\sim3$ $\mu$m in the outer disk. In this case, the scattered light has insufficient optical depth to be detected at the infrared wavelengths. 
According to our study with millimeter polarization, the dust scale height is constrained to be higher outside the 70 au ring than inside the 70 au ring.
This is consistent with both models.
For the first model, because there is no constraint on the dust scale height of the 70 au ring itself, the 70 au ring may have a dust scale height that is as large as the gas scale height.
The dust scale height was derived to be two-thirds the gas scale height outside the 70 au ring, and thus the outer regions will be in the shadow cast by the 70 au ring.
The second model is also consistent with our dust scale height model because there is no constraint on the dust scale height of the small dust grains from the SPHERE observations.
It would be possible to distinguish the two models by performing millimeter-wave polarization observations at wavelengths longer than 0.87 mm to detect the self-scattering emission from the rings to constrain the dust scale height of the 70 au ring.

Observations of the disk were also carried out with {\it HST} \citep{gra00,wis08}.
Dust  in  the  disk  was  detected  in  scattered  light out to $\sim500$ au, which indicates that the outermost  regions  are  flaring  and small dust grains are directly  illuminated  by  the central  star.
\citet{wis08} found that the disk color ($V-I$) becomes redder at a radial distance of $3\farcs5$ corresponding to 355 au.
They suggested that the red disk color shows evolution in the grain size distribution such as grain growth.
This may be consistent with the depletion of the smallest dust grains in the second model proposed by \citet{mur18}.

\subsection{Caveats of models}
In Section 5.2 and 5.3, we used two populations of dust grains to reproduce the observations. The maximum grain sizes were set to $140$ $\mu$m and 1 mm, respectively.
Even though we set the 1 mm dust grains in the rings, it is possible that dust grains much larger or smaller than 140 $\mu$m are distributed in the rings rather than the 1 mm dust grain population.
To further constrain the dust size in the rings, polarization observation at longer or shorter wavelengths are needed.

Another caveat is that the polarization is assumed to be produced only by self-scattering.
We show that the polarization of this disk at a 0.87 mm wavelength can be understood only by self-scattering.
However, polarization can also be produced by grain alignment mechanisms such as magnetic fields, radiation gradients, and gas flows.
The observed polarization may also be contributed from the thermal emission of aligned dust grains. In this case, the azimuthal component may be explained by the dust alignment rather than the dust scale height of self-scattering.
To investigate the contribution to polarization from dust alignment, higher spatial resolution and different wavelength observations are needed.

\section{Conclusion} \label{sec:conclusion}

The protoplanetary disk around HD 163296 shows polarized emission at a 0.87 mm wavelength \citep{den19}, where ring and gap structures are spatially resolved.
Motivated by the spatially resolved polarization of the ring-and-gap disk, we performed radiative transfer calculations to determine the key parameters to calculate the polarization vectors and fraction in a disk with rings and gaps.
Furthermore, we also developed a model to reproduce the polarization of HD 163296 to constrain the disk parameters.
The main findings are summarized as follows.
\begin{itemize}
\item The key parameters that determine the polarization vectors and fraction due to self-scattering in an inclined disk are the dust scale height and the dust grain size or population, as summarized in Figure \ref{fig1}.
As the dust scale height increases, the polarization fraction increases more along the minor axis than along the major axis.
As a result, the azimuthal variation of the polarization fraction is enhanced.
To further investigate the effects of grain size, we used two populations of dust grains, with maximum grain sizes of 140 $\mu$m and 1 mm, respectively.
As the ratio of the 1 mm grain population increases, the polarization fraction decreases.
This is because the 140 $\mu$m population produces polarization whereas the 1 mm population does not.

\item From the radiative transfer modeling of HD 163296 polarization observations, we found that the gaps require maximum efficiency of the polarization whereas the rings and the central region do not.
In terms of grain size, the gaps are likely to be filled with the 140 $\mu$m grain population and the central region and rings are likely to have a significant contribution from a dust grain population with a grain size of much larger or smaller than 140 $\mu$m.
The rings are consistent with no contribution from the 140 $\mu$m population.

\item We compared the size constraints with the spectral index analysis of HD 163296.
The low spectral index ($\alpha\sim1.5-2.0$) in the central part of the disk may indicate optically thick emission that is further reduced by dust scattering with a size of $\sim100$ $\mu$m. The low spectral index ($\alpha\sim2.0$) in the 70 au ring may be caused by the large optical depth and/or grain growth effects because the 70 au ring is optically thick ($\tau\sim1.1$).

\item The model indicates that the dust scale height is less than (1/3) $H_{\rm gas}$ inside the 70 au ring and as high as $(2/3) H_{\rm gas}$ outside the 70 au ring to explain the Stokes {\it Q} image.
This suggests that the turbulence strength differ between the regions inside and outside the 70 au ring.
We estimated the gas turbulence parameter at the two gaps after constraining the grain size and the dust scale height. The turbulence $\alpha$ is $\alpha\lesssim1.5\times10^{-3}$ in the gap inside the 70 au ring and $\alpha\sim0.015-0.3$ in the gap outside the 70 au ring. 
A dead zone or e-heating zone may explain this transition.
The dead zone scenario is qualitatively consistent with dust accumulation at the outer edge of the dead zone and ring formation.
\end{itemize}

The authors thank the anonymous referee, whose comments led to substantial improvements in the manuscript.
The authors also thank S. Mori and T. Ueda for fruitful discussion and comments.
SO thanks W.R.F. Dent, C.L.H. Hull, and S. Ishii. They hosted SO in  the Joint ALMA Observatory, Chile. SO acknowledges support from the Joint ALMA Observatory Visitor Program.
This work was supported by JSPS KAKENHI Grant Numbers 18K13590 and 18K13595. 
This paper makes use of the following ALMA data: ADS/JAO.ALMA\#2015.1.00616.S. ALMA  is a partnership of ESO (representing its member states), NSF (USA) and NINS (Japan),  together with NRC (Canada), NSC and ASIAA (Taiwan), and KASI (Republic of Korea), in  cooperation with the Republic of Chile. The Joint ALMA Observatory is operated by  ESO, AUI/NRAO and NAOJ.
Data analysis was in part carried out on the Multi-wavelength Data Analysis System operated by the Astronomy Data Center (ADC), National Astronomical Observatory of Japan.

\facilities{ALMA}

\software{CASA \citep[v4.5.3;][]{mcm07}, RADMC-3D \citep{dul12}
          }




\appendix

\section{Additional Test of Model with Single Grain Population}\label{ad_test}
\begin{figure}[htbp]
  \begin{center}
  \includegraphics[width=16cm,bb=0 0 1594 857]{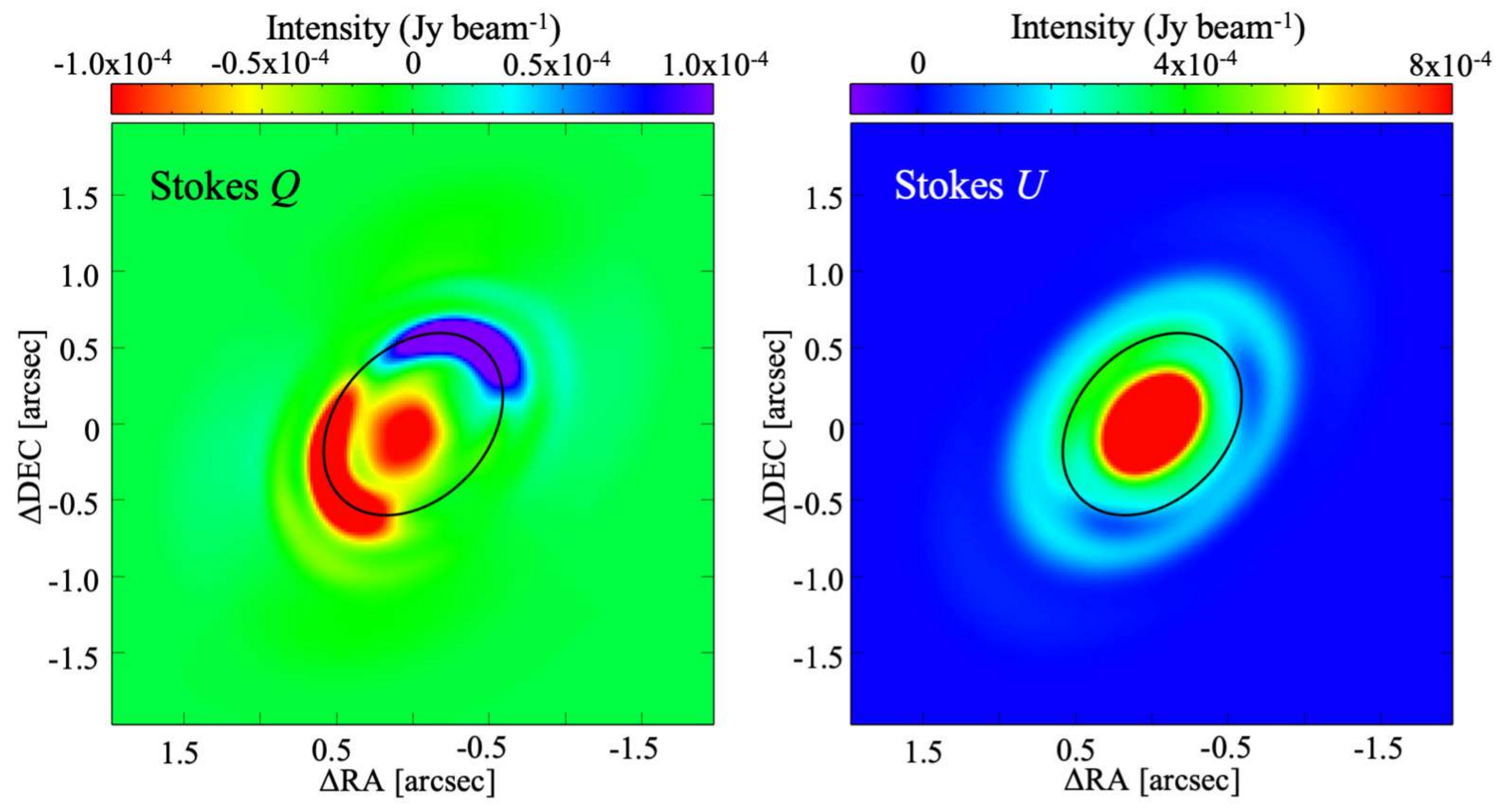}
    \end{center}
  \caption{Stokes {\it Q} and {\it U} images of the model with a single grain population. Dust grains with a size of 140 $\mu$m are distributed in both the rings and gaps. The dust scale height ranges from ${\rm fset}=1.5$ to ${\rm fset=10}$ at 68 au.
  }
  \label{single_grain_fset}
\end{figure}

In Section 5.1, we suggested that a model with a single grain population cannot reproduce the observations with the assumption of a radially constant dust settling parameter $f_{\rm set}$.
Here, we investigate a model with a single grain population that takes into account the radial variation of the dust scale height.
We compared the observations with the model with a single grain population by changing the dust scale height near the 70 au ring. We used a single grain population with a grain size of $140$ $\mu$m, which is the same model as that described in Section 5.1. We applied $f_{\rm set}=10$ within $r\leq68$ au and $f_{\rm set}=1.5$ for $r>68$ au.  We found that the Stokes {\it Q} emission is produced around the 70 au ring, as shown in Figure \ref{single_grain_fset}, because the dust grains that contribute to the polarization are located on the ring. However, the observations show that the Stokes {\it Q} emission is not detected on the 70 au ring.
Therefore, a model with a single grain population is unlikely to explain the observations even if the dust scale height is changed.

\section{Original Images of Best Model and Future Prospects}\label{original_image}
Here, we discuss possible future observations with high spatial resolutions by showing the original image before convolution with the beam size.
As shown in Figures \ref{comparison} and \ref{model_best_plot}, our best model well matches the observations.

\begin{figure}[htbp]
  \begin{center}
  \includegraphics[width=16cm,bb=0 0 2880 1687]{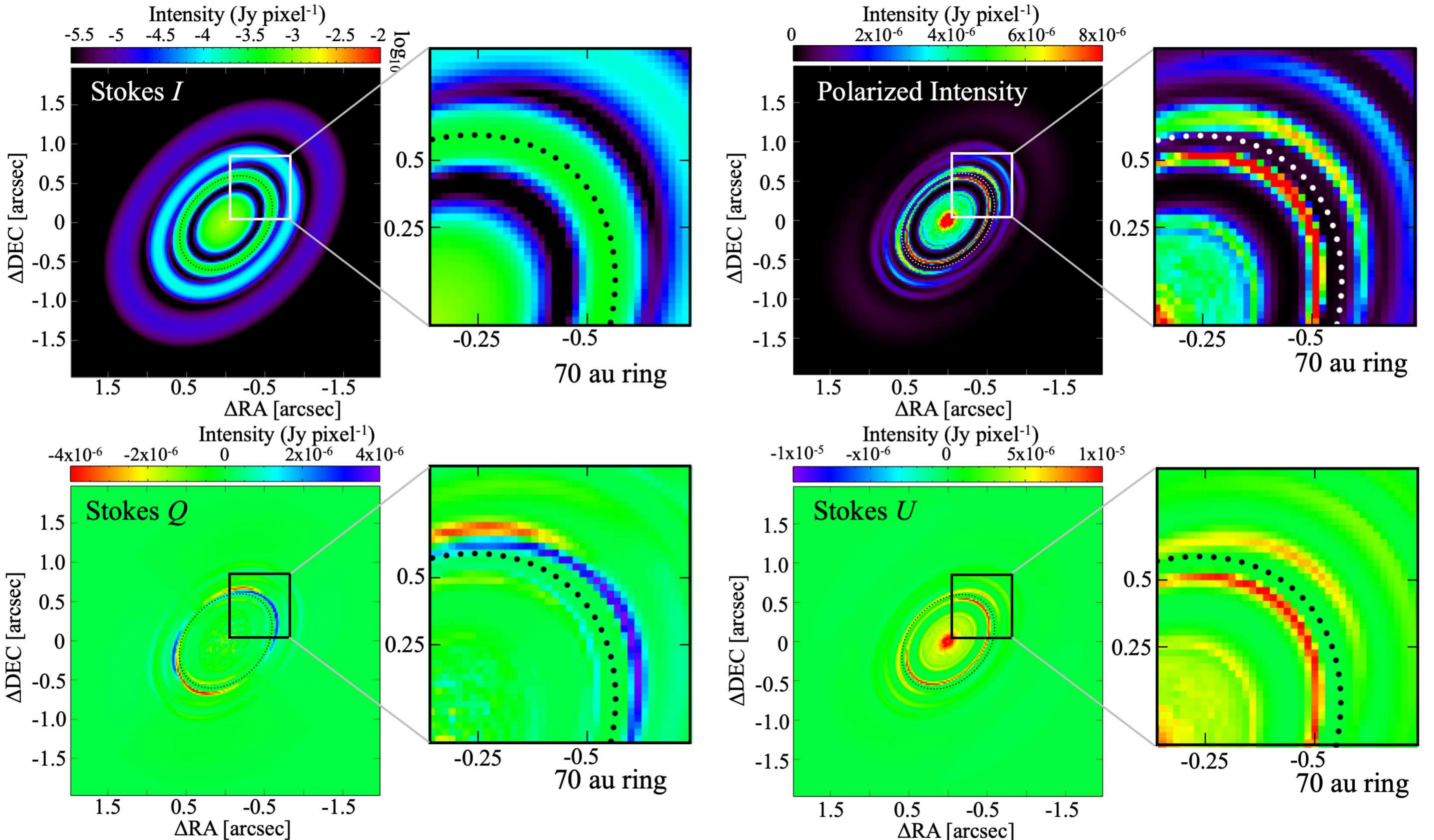}
    \end{center}
  \caption{Original images of the Stokes {\it I}, {\it Q}, and {\it U} and the polarized intensity of our best model. The black and white dots show the 70 au ring. These images are not smoothed by the Gaussian beam.  The unit of the intensity is set to Jy pixel$^{-1}$. The pixel size is $0\farcs02$.
  }
  \label{original}
\end{figure}

The original images of the Stokes {\it I}, {\it Q}, and {\it U} and the polarized intensity of the model are shown in Figure \ref{original}. These images are not smoothed by the Gaussian beam.
We predict that the emission of the Stokes {\it Q} and {\it U} and the polarized intensity cannot be detected in the rings because dust grains larger or smaller than 140 $\mu$m cannot produce polarization due to self-scattering at a 0.87 mm wavelength. The peak emission of the Stokes {\it U} and the polarized intensity could be detected inside and outside the 70 au ring. Observations with a large beam size thus show peak emission of the Stokes {\it U} and the polarized intensity around the 70 au ring.
The Stokes {\it Q} emission could be detected only outside the 70 au ring because it is not produced inside the 70 au ring due to the low dust scale height ($H_{\rm dust}\lesssim0.33 H_{\rm gas}$).
We can constrain the dust scale height inside the 70 au ring if we detect the Stokes {\it Q} emission in the future with higher-sensitively observations (see Section 5.4).



\end{document}